\documentclass[useAMS,referee,usenatbib]{biom}
\usepackage{natbib}
\usepackage{url}
\usepackage{bm}
\usepackage[flushleft]{threeparttable}
\usepackage{booktabs}
\usepackage{rotating}
\usepackage{adjustbox}
\usepackage{xcolor}
\usepackage{changepage}
\usepackage{verbatim}
\usepackage{amsmath}
\usepackage{amsfonts}
\usepackage{graphicx}
\usepackage{tikz}
\usepackage{multirow}
\usepackage{subcaption}
\usepackage{bigints}

\usepackage{hyperref}
\hypersetup{colorlinks=true, linkcolor=blue, citecolor=blue, filecolor=blue}

\usepackage{enumerate}
\usepackage[medium,compact]{titlesec}
\allowdisplaybreaks

\def\bSig\mathbf{\Sigma}

\title[Marginal structural quantile model]{Doubly robust estimation and sensitivity analysis for marginal structural quantile models}

\author{Chao Cheng$^{1,2,*}$\email{c.cheng@yale.edu}, 
Liangyuan Hu$^{3}$, and Fan Li$^{1,2,**}$\email{fan.f.li@yale.edu} \\
$^{1}$Department of Biostatistics, Yale School of Public Health, New Haven, CT \\
$^{2}$Center for Methods in Implementation and Prevention Science, Yale School of Public Health, New Haven, CT \\
$^{3}$Department of Biostatistics and Epidemiology, Rutgers University, New Brunswick, NJ, USA}

\begin{document}





\pagerange{\pageref{firstpage}--\pageref{lastpage}} 
\volume{64}
\pubyear{2008}
\artmonth{December}


\doi{10.1111/j.1541-0420.2005.00454.x}


\label{firstpage}


\begin{abstract}
The marginal structure quantile model (MSQM) provides a unique lens to understand the causal effect of a time-varying treatment on the full distribution of potential outcomes. Under the semiparametric framework, we derive the efficiency influence function for the MSQM, from which a new doubly robust estimator is proposed for point estimation and inference. We show that the doubly robust estimator is consistent if either of the models associated with treatment assignment or the potential outcome distributions is correctly specified, and is semiparametric efficient if both models are correct. To implement the doubly robust MSQM estimator, we propose to solve a smoothed estimating equation to facilitate efficient computation of the point and variance estimates. In addition, we develop a confounding function approach to investigate the sensitivity of several MSQM estimators when the sequential ignorability assumption is violated. Extensive simulations are conducted to examine the finite-sample performance characteristics of the proposed methods. We apply the proposed methods to the Yale New Haven Health System Electronic Health Record data to study the effect of antihypertensive medications to patients with severe hypertension and assess the robustness of findings to unmeasured baseline and time-varying confounding.
\end{abstract}

%



\maketitle


%


\section{Introduction}\label{sec:intro}

In longitudinal observational studies, the causal effect of a time-varying treatment is typically defined based on the mean potential outcome. However, in health sciences research, outcomes often exhibit heavy tails or skewness. In such cases, the quantiles of these outcomes can provide a more meaningful summary measure. For example, the tail of the outcome distribution is useful for medical decision-making, when we assess the effect of antihypertensives on blood pressure among patients who developed severe inpatient hypertension \citep{ghazi2022severe}. In this context, the causal effects defined on the upper quantiles of the blood pressure distribution are relevant for safety considerations, whereas the average causal effect defined on the mean potential outcome does not adequately capture shifts in the extremes.

With a static time-varying treatment, there have been limited methods for estimating the quantile causal effects. \cite{robins2000marginal} developed the structural nested distribution models (SNDMs) and the method of $g$-estimation for evaluating the longitudinal treatment effect on the entire distribution of the outcome. However, the resulting estimator is generally computationally intensive with a large number of time points and limited software is available to implement SNDMs. {\color{black}As a computationally efficient alternative, \cite{hogan2004marginal} introduced the marginal structural quantile model (MSQM) and an accompanying inverse probability weighting (IPW) estimator that requires specification of a series of parametric working models on the longitudinal propensity scores. While the IPW estimation of MSQM is straightforward to implement, its consistency relies on the correctly specified parametric propensity score models. Although the efficiency of IPW can be improved by using a nonparamteric propensity score estimator with a point treatment \citep{ertefaie2023nonparametric}, IPW based on parametric propensity scores frequently leads to an inefficient estimator with either a point treatment and a time-varying treatment. To date, the development of a more efficient MSQM estimator that improves upon IPW alone remains unexplored.} 

In this article, we develop a semiparametric framework to obtain new estimators for identifying the causal effects of a time-varying treatment parameterized by the MSQM. As a complement to the IPW, we first propose an iterative conditional regression (ICR) approach based on a sequence of nested regressions for the outcome distributions given the time-dependent covariates and treatments. {\color{black}Inspired from the existing efficiency theory of the marginal structural mean model (MSMM) \citep{van2003unified}, we derive the efficient influence function (EIF) for the causal parameters in the MSQM, which motivates a doubly robust and locally efficient estimator that combines IPW and ICR. Similar to the existing doubly robust MSMM estimator (\citealp{van2003unified,bang2005doubly}), our doubly robust MSQM estimator offers stronger protection against model misspecification, and permits consistent estimation of when either the models associated with IPW or those with ICR are correctly specified, but not necessarily both. However, construction of the doubly robust MSQM estimator differs from the doubly robust MSMM estimators due to its two distinct features: (1) the EIF of MSQM  is discontinuous in the causal parameters and (2) the EIF of MSQM involves the full potential outcome distribution as a nuisance, which necessitates the specification of the entire potential outcome distribution rather than just the mean as in MSMM.} To alleviate the computational challenge with discontinuity, we consider a local distribution function to construct an asymptotically equivalent and yet smoothed estimating equation; this permits the use of a standard derivative-based algorithm for point estimation as well as fast computation of the asymptotic standard errors. Finally, our estimators for MSQM require the sequential ignorability assumption \citep{robins1999estimation}, which assumes the absence of unmeasured baseline and time-dependent confounding. Because the ignorability assumption is not empirically verifiable from the observed data alone, we additionally contribute a new sensitivity analysis strategy to assess the impact on MSQM estimates in the presence of unmeasured confounding.

\section{Marginal Structural Quantile Model} 
\label{sec:msqm}
We consider a longitudinal observational study with $n$ individuals that may receive treatments during $K$ periods. For $k=1,\dots,K$, let $A_k$ be the treatment administered during the $k$-th period that takes values in a finite set $\mathbb{A}_k$ and $\bm L_k$ be a vector of time-varying confounders measured at the beginning of the $k$-th period. We consider $\mathbb{A}_k=\{0,1\}$ such that $A_k=1$ indicates treatment and $A_k=0$ indicates control. We assume that a final outcome $Y$ is observed at the end of the $K$-th period and is of scientific interest. To summarize, the observed data consists of $n$ i.i.d. realizations of $\bm O = \{\bm L_1, A_1,\bm L_2, A_2,\dots, \bm L_{K},A_K, Y\}$. Throughout, we use the overbar/underbar notation to denote all past/future values of a variable; for example, $\overline A_k = (A_1,\dots,A_k)$ is the treatment history up to the $k$-th period and $\underline A_k = (A_k,\dots,A_K)$ is the treatment history from the $k$-th period onward to the $K$-th period. Let $\overline{\mathbb A}_k = \{0,1\}^{\otimes k}$ denote all possible treatment assignments in the first $k$ periods; $\underline{\mathbb A}_k = \{0,1\}^{\otimes(K-k+1)}$ is similarly defined. We use $f_{Z|X}(z|x)$ and $F_{Z|X}(z|x)$ to denote the density function and the cumulative distribution function (CDF) of $Z$ given $X$, respectively. 

Corresponding to each static treatment regimen $\overline a_K \in \overline{\mathbb A}_K$, let $Y_{\overline a_K}$ be the potential outcome had the unit followed treatment regimen $\overline a_K$. We assume $Y_{\overline a_K}$ is continuously distributed with support $[y_{\min},y_{\max}]\in\mathbb{R}$, where $y_{\min}$ and $y_{\max}$ can be $-\infty$ and $\infty$. We specify the following MSQM for $Y_{\overline a_K}$ with unknown causal parameters $\bm \theta_q \in \mathbb{R}^p$: 
\begin{equation}\label{msqm}
    Q_{Y_{\overline a_K}|\bm Z}^{(q)} = h(\overline a_K, \bm Z;\bm \theta_q) \quad \forall \overline a_K \in \overline{\mathbb{A}}_K,
\end{equation}
where $Q_{Y_{\overline a_K}|\bm Z}^{(q)} = \text{inf}\{y: F_{Y_{\overline a_K}|\bm Z}(y|\bm Z) \geq q\}$ is the $q$-th quantile of $F_{Y_{\overline a_K}|\bm Z}(y|\bm Z)$, $h$ is a pre-specified smooth function, and $\bm Z$ is a subset of baseline covariates $\bm L_1$. The MSQM \eqref{msqm} is a model of science that measures how quantiles of the potential outcome varies as a function of the treatment regimen and possibly baseline covariates. For example, one may assume $h(\overline a_K, \bm Z;\bm \theta_q)=\theta_{0q}+\sum_{k=1}^K \theta_{kq}a_k + \bm \theta_{zq}^T \bm Z$, where $\bm \theta_q = [\theta_{0q},\theta_{1q},\dots, \theta_{Kq},\bm \theta_{zq}^T]^T$. Then the treatment regimens, $\overline a_K$ versus $\overline a_K^{'}$, can be compared by contrasting $Q_{Y_{\overline a_K}|\bm Z}^{(q)}$ versus $Q_{Y_{\overline a_K^{'}}|\bm Z}^{(q)}$ for a specific $q\in (0,1)$. In practice, 
$
\tau_{\overline a_K,\overline a_K^{'}|\bm Z}^{(q)} = h(\overline a_K, \bm Z;\bm \theta_q) - h(\overline a_K^{'}, \bm Z;\bm \theta_q)
$
can be used to measure the quantile causal effect by switching the treatment regimen from $\overline a_K^{'}$ to $\overline a_K$.

Our interest lies in estimating the structural parameter $\bm\theta_q$ in model \eqref{msqm}. We define the causal parameter of interest as the solution of the following equation with respect to $\bm\theta_q$:
\begin{equation}\label{msqm_equation}
E\left[\sum_{\overline a_K \in \overline{\mathbb A}_K} d(\overline a_K,\bm Z;\bm\theta_{q})\Big\{\mathbb{I}\left(Y_{\overline a_K}\leq h(\overline a_K, \bm Z;\bm \theta_{q})\right)- q\Big\}  \right] = \bm 0,
\end{equation}
where $d(\overline a_K,\bm Z;\bm\theta_{q}) = \rho(\overline a_K,\bm Z) \times \frac{\partial}{\partial \bm \theta_q} h(\overline A_K, \bm Z;\bm \theta_q)$ is a $p\times 1$ matrix and $\rho(\overline a_K,\bm Z)$ is a weight function that can depend on $\overline a_K$ and $\bm Z$; a common choice is $\rho(\overline a_K,\bm Z) = 1$. {\color{black}With a more general choice of $\rho(\overline a_K,\bm Z)$, we leverage insights from \citet{angrist2006quantile} to show that $\bm\theta_q$ in \eqref{msqm_equation}  embodies a nonparametric causal effect in the vein of \citet{neugebauer2007nonparametric}. This interpretation is valid in that $\bm\theta_{q}$ accurately parameterizes the true quantile causal curve when the MSQM is correctly specified but remains a localized and approximate summary measure of the quantile causal curve when the MSQM is misspecified (Web Appendix A).} Up until Section \ref{sec:dr}, we consider the following identifying assumptions: (A1) \textit{consistency}, $Y_{\overline a_K} = Y$ if $\overline A_K = \overline a_K$; (A2) \textit{sequential ignorability}, for $k=1,\dots,K$, $Y_{\overline a_K} \perp A_k | \{\overline{\bm L}_k, \overline A_{k-1}=\overline a_{k-1}\}$, $\forall ~\overline a_K \in \overline{\mathbb{A}}_K$; and (A3) \textit{positivity}, for all $k$ and $\overline a_K$, if $f_{\overline A_{k-1},\overline{\bm L}_k}(\overline a_{k-1},\overline{\bm l}_{k-1})>0$, then $f_{A_k|\overline A_{k-1}, \overline{\bm L}_{k}}(a_k|\overline a_{k-1},\overline{\bm l}_k)>0$. In addition, we require (A4) \textit{uniqueness}, the solution of \eqref{msqm_equation} in terms of $\bm\theta_q$ is unique. 
The sequential ignorability assumption (A2) is empirically unverifiable, and a sensitivity analysis framework for Assumption (A2) will be addressed in Section \ref{sec:sensitivity}.

\section{Inverse Probability Weighting and Iterative Conditional Regression}
\label{sec:ipw-icr}

\subsection{IPW estimation: a brief review}\label{sec:IPW-alone}

For $k=1,\dots,K$, let $\pi_k(\overline a_k,\overline{\bm l}_k) = f_{A_k|\overline A_{k-1},\overline{\bm L}_k}(a_k|\overline a_{k-1},\overline{\bm l}_k)$ be the probability mass function of $A_k=a_k$ conditional on $\overline A_{k-1}=\overline a_{k-1}$ and $\overline{\bm L}_k=\overline{\bm l}_k$, which is referred to as the propensity score. 
We can posit a parametric model $\pi_k(\overline a_k, \overline{\bm l}_k; \bm \alpha_k)$ with parameter $\bm \alpha_k$, and $\widehat{\bm \alpha}_k$ can be obtained by solving the score equation $ \mathbb{P}_n \left[\frac{\partial}{\partial \bm \alpha_k} \log \pi_{k}(\overline A_{k}, \overline{\bm L}_k;\bm \alpha_k) \right]= \bm 0$. Here, $\mathbb{P}_n[\cdot] = \frac{1}{n} \sum_{i=1}^n [\cdot]$ defines an empirical average. Under assumptions (A1)--(A4), equation \eqref{msqm_equation} is equivalent to  
$
E\left[\frac{d(\overline A_K,\bm Z;\bm\theta_q)}{\overline \pi_K(\overline A_K,\overline{\bm L}_K)}\left\{\mathbb{I}\Big(Y\leq h(\overline A_K, \bm Z;\bm \theta_q) \Big) - q\right\}\right] = \bm 0,
$
where $\mathbb{I}(\cdot)$ is the indicator function and $\overline \pi_k(\overline A_k,\overline{\bm L}_k) = \prod_{j=1}^k \pi_j(\overline A_j,\overline{\bm L}_j)$ is the cumulative probability to receive $\overline A_k$ in the first $k$ periods. The IPW estimator (\citealp{hogan2004marginal}), denoted by $\widehat{\bm \theta}_q^{\text{IPW}*}$, is obtained by solving
\begin{equation}\label{ee:ipw}
\mathbb{P}_n \left[ \frac{d(\overline A_K,\bm Z;\bm\theta_q)}{ \overline{\pi}_K(\overline A_K, \overline{\bm L}_K;\widehat{\bm\alpha})} \left\{\mathbb{I}\Big(Y \leq h(\overline A_K, \bm Z;\bm \theta_q)\Big)-q\right\}\right] = \bm 0,
\end{equation}
where $\widehat{\bm\alpha}=[\widehat{\bm\alpha}_1^T,\dots,\widehat{\bm\alpha}_K^T]^T$ and $\overline{\pi}_K(\overline A_K, \overline{\bm L}_K,\widehat{\bm\alpha})=\prod_{k=1}^K \pi_k(\overline A_k,\overline{\bm L}_k;\widehat{\bm \alpha}_k)$ is the inverse probability of treatment weight. Let $\mathcal{M}_{ps}$ be the sequence of propensity score models $\pi_k(\overline a_k,\overline{\bm l}_k;\bm\alpha_k)$, $k=1,\dots,K$. \cite{hogan2004marginal} showed $\widehat{\bm \theta}_q^{\text{IPW}*}$ is consistent if $\mathcal M_{ps}$ is correctly specified.

\subsection{Iterative conditional regression (ICR)}\label{sec:ICR-alone}

We next develop an alternative estimator for $\bm\theta_q$ by modeling the potential outcome distribution. Using the law of iterated expection, 
we can show that \eqref{msqm_equation} is equivalent to
\begin{equation}\label{icr-t0}
E\left[\sum_{\overline a_K \in \overline{\mathbb A}_K} d(\overline a_K,\bm Z;\bm\theta_{q})\left\{F_{Y_{\overline a_K}|\bm L_1}\left(h(\overline a_K, \bm Z;\bm \theta_{q})\Big|\bm L_1\right)- q\right\}  \right] = \bm 0.
\end{equation} 
Therefore, estimating $\bm\theta_q$ boils down to estimating $F_{Y_{\overline a_K}|\bm L_1}$, or equivalently $f_{Y_{\overline a_K}|\bm L_1}$. Typically, one can estimate $f_{Y_{\overline a_K}|\bm L_1}$ by first postulating parametric models for $f_{Y|\overline A_K, \overline{\bm L}_K}$ and $f_{\bm L_k |\overline A_{k-1}, \overline{\bm L}_{k-1}}$, $k=2,\dots,K$, and then computing $f_{Y_{\overline a_K}|\bm L_1}$ based on  \cite{robins2000marginal}'s $g$-formula,
$
f_{Y_{\overline a_K}|\bm L_1}\left(y|\bm l_1\right) = \int_{\bm l_{K}} \dots \int_{\bm l_{2}} f_{Y|\overline A_K, \overline{\bm L}_K}(y|\overline a_K, \overline{\bm l}_K)\times \prod_{k=2}^K f_{\bm L_k |\overline A_{k-1}, \overline{\bm L}_{k-1}}(\bm l_k|\overline a_{k-1},\overline{\bm l}_{k-1}) \text{d}\bm l_k
$ (Lemma 1 in the Supplementary Material). However, this may be cumbersome due to the need to model all time-varying covariates, $\bm L_k$, which are usually multi-dimensional. We consider an alternative procedure to estimate $f_{Y_{\overline a_K}|\bm L_1}$ without modeling the time-varying covariates. 

For $k=1,\dots,K$, let $f_{Y_{\overline A_k,\underline a_{k+1}}|\overline A_k,\overline{\bm L}_k}$ be the the  density of $Y_{\overline A_k,\underline a_{k+1}}$ given $\overline A_k$ and $\overline{\bm L}_k$. Here, for notational convenience, we let $Y_{\overline A_k,\underline{a}_{k+1}}$ be the potential outcome had the unit been treated with the observed treatment history $\overline A_k$ in the first $k$ time periods and $\underline{a}_{k+1}$, rather than the observed $\underline{A}_{k+1}$, from time $k+1$ to the end of study. It is immediate that $Y_{\overline A_k,\underline{a}_{k+1}} = Y_{\overline a_K}$ if $\overline A_k = \overline a_k$ and therefore $f_{Y_{\overline A_k,\underline a_{k+1}}|\overline A_k,\overline{\bm L}_k}(y|\overline a_k,\overline{\bm l}_k)=f_{Y_{\overline a_K}|\overline A_k,\overline{\bm L}_k}(y|\overline a_k,\overline{\bm l}_k)$. Our goal is to estimate $f_{Y_{A_1,\underline a_2}|A_1,\bm L_1}(y|a_1,\bm l_1)$, which is identical to $f_{Y_{\overline a_K}|\bm L_1}(y|\bm l_1)$ by Assumption (A2), and $f_{Y_{\overline a_K}|\bm L_1}(y|\bm l_1)$ is the key density in \eqref{icr-t0} used to solve for $\bm\theta_q$. In practice, one posit a parametric model $\psi_k(Y_{\overline A_k,\underline{a}_{k+1}},\overline A_k,\underline a_{k+1},\overline{\bm L}_k;\bm \beta_k)$ with parameter $\bm \beta_k$ for the density $f_{Y_{\overline A_{k},\underline a_{k+1}}|\overline A_k,\overline{\bm L}_k}$. 

If $Y_{\overline A_k,\underline{a}_{k+1}}$ for all $\underline{a}_{k+1} \in \underline{\mathbb A}_{k+1}$ is fully observed, then $\bm\beta_k$ can be estimated by solving
\begin{equation}\label{ee-icr:complete}
    \mathbb{P}_n \left[ \sum_{\underline{a}_{k+1} \in \underline{\mathbb A}_{k+1} } \mathbb{U}_{\bm \beta_k}(Y_{\overline A_k, \underline{a}_{k+1}}, \overline A_k, \underline{a}_{k+1}, \overline{\bm L}_k)\right]=\bm 0.
\end{equation}
Here, $\mathbb{U}_{\bm \beta_k}(Y_{\overline A_k, \underline{a}_{k+1}}, \overline A_k, \underline{a}_{k+1}, \overline{\bm L}_k)$ is an unbiased estimating function; an example is the score function $\mathbb{U}_{\bm \beta_k}(Y_{\overline A_k, \underline{a}_{k+1}}, \overline A_k, \underline{a}_{k+1},\overline{\bm L}_k) = \frac{\partial }{\partial \bm \beta_k} \log \psi_k(Y_{\overline A_{k},\underline a_{k+1}},\overline A_{k}, \underline{a}_{k+1},\overline{\bm L}_k;\bm \beta_k)$. When $k=K$, the estimating equation \eqref{ee-icr:complete} degenerates to $\mathbb{P}_n \left[ \mathbb{U}_{\bm \beta_K}(Y, \overline A_K, \overline{\bm L}_K)\right]= \bm 0$ since $Y_{\overline A_K}=Y$ by Assumption (A1), and therefore $\widehat{\bm\beta}_K$ can be obtained by directly solving \eqref{ee-icr:complete}. When $k<K$, $Y_{\overline A_k, \underline{a}_{k+1}}$ for $\underline{a}_{k+1} \neq \underline{A}_{k+1}$ are unobserved, therefore $\bm\beta_k$ can not be estimated through \eqref{ee-icr:complete}. We leverage the idea of the expected estimating equation (\citealp{wang2000expected}) to address this structural missing data problem. We construct an estimating equation as the expectation of the complete-data estimating equation \eqref{ee-icr:complete} with respect of a known distribution of the missing variable $Y_{\overline A_k,\underline{a}_{k+1}}$, which depends only on the observed data. Specifically, if $\bm\beta_{k+1}$ is known, we estimate $\bm\beta_k$ by solving
\begin{equation}\label{icr2}
\mathbb{P}_n\left[\sum_{\underline{a}_{k+1} \in \underline{\mathbb A}_{k+1} }\int_y \mathbb U_{\bm \beta_k}(y, \overline A_k, \underline{a}_{k+1}, \overline{\bm L}_k) \psi_{k+1}(y,\overline A_k, \underline{a}_{k+1}, \overline{\bm L}_{k+1};\bm\beta_{k+1}) \text{d} y \right] = \bm 0,
\end{equation}
which is the expectation of \eqref{ee-icr:complete} with respect of $\psi_{k+1}(y,\overline A_k, \underline{a}_{k+1}, \overline{\bm L}_{k+1};\bm\beta_{k+1})$; i.e., the conditional distribution of $Y_{\overline A_k, \underline{a}_{k+1}}$ given that the individual received $(\overline A_k,a_{k+1})$ and observed $\overline{\bm L}_{k+1}$. We show in Web Appendix B that \eqref{icr2} is an unbiased estimating equation. 

The expected estimating equation suggests the following iterative procedure to estimate $\bm \beta_k$ from $k=K,\dots,1$. First, we obtain $\bm\beta_K$ by solving $\mathbb{P}_n \left[ \mathbb{U}_{\bm \beta_K}(Y, \overline A_K, \overline{\bm L}_K)\right]= \bm 0$; then, for $k=K-1,\dots,1$, we iteratively compute $\widehat{\bm\beta}_k$ by solving \eqref{icr2} with $\bm\beta_{k+1}$ fixed at $\widehat{\bm \beta}_{k+1}$. Finally, we solve the following estimating equation invoked by the restriction \eqref{icr-t0} to obtain $\bm \theta_q$,
\begin{equation}\label{ee:icr-t0}
    \mathbb{P}_n\left[\sum_{\overline a_K \in \overline{\mathbb{A}}_K} d(\overline a_K,\bm Z;\bm\theta_q)\left\{\Psi_1\left(h(\overline a_K, \bm Z;\bm \theta_q),\overline a_K,\bm L_1;\widehat{\bm\beta}_1\right) - q\right\} \right] = \bm 0,
\end{equation}
where  $\Psi_k(y,\overline a_K,\overline{\bm l}_k;\bm \beta_k) \equiv \int_{y_{\min}}^{y}  \psi_k(u,\overline a_K,\overline{\bm l}_k;\bm \beta_k) \text{d}u$ is the CDF corresponding to $\psi_k(u,\overline a_K,\overline{\bm l}_k;\bm \beta_k)$, and is equal to $F_{Y_{\overline A_k,\underline{a}_{k+1}}|\overline A_k,\overline{\bm L}_k}(y|\overline a_k,\overline{\bm l}_k)$ when evaluated at the true $\bm\beta_k$. We shall denote the estimator given by this iterative procedure by $\widehat{\bm \theta}_q^{\text{ICR}}$.  Let $\mathcal{M}_{om}$ be the sequence of outcome regressions $\psi_k(Y_{\overline A_k,\underline{a}_{k+1}},\overline A_k,\underline a_{k+1},\overline{\bm L}_k;\bm \beta_k)$, $k=1,\dots,K$. In Web Appendix B.1, we show that $\widehat{\bm \theta}_q^{\text{ICR}}$ is consistent and asymptotic normal (CAN) if $\mathcal{M}_{om}$ is correctly specified. A consistent variance estimator is also provided for inference. In Web Appendix B.2, we provide a concrete example of the ICR approach under nested Gaussian linear regression, allowing for possible outcome transformation and heteroscedasticity.

{\color{black}In Web Appendix B.2, we provide clarification that the ICR approach expands on the iterative conditional expectation (ICE) estimator used for MSMM \citep{bang2005doubly}. Both approaches  depend only outcome modeling for inference about the structural parameters. The key distinction between them is that the ICR approach exploits the full outcome distribution for estimating MSQM, whereas the ICE approach is contingent on modeling the mean outcome for estimating MSMM. A direct byproduct of the ICR approach is the estimate of $\{F_{Y_{\overline A_k,\underline a_{k+1}}|\overline A_k,\overline{\bm L}_k},k=1,\dots,K\}$, a sequence of nuisance functions that are involved in constructing the doubly robust estimator in Section \ref{sec:dr}. Therefore, the ICR approach allows us to incorporate the unknown nuisance functions into the doubly robust estimator for MSQM, which parallels the strategy in constructing doubly robust MSMM estimators \citep{van2003unified,bang2005doubly}.}

\section{Doubly Robust Estimation and Inference}\label{sec:dr}



Validity of IPW or ICR requires correct specification of $\mathcal M_{ps}$ or $\mathcal M_{om}$, respectively. To improve upon such singly robust estimators, we propose a doubly robust estimator of $\bm\theta_q$. The development of a doubly robust MSQM estimator relies on the efficient influence function (EIF) of our target estimand $\bm\theta_{q}$ (\citealp{bickel1993efficient}). Moreover, the semiparametric efficiency lower bound of $\bm\theta_q$ is determined by the variance of the EIF. We provide the explicit form of the EIF for $\bm\theta_{q}$ in Theorem \ref{EIF} to motivate our new doubly robust estimator. 

\begin{theorem}\label{EIF}
The EIF for $\bm\theta_{q}$ is $\mathbb{U}_{\bm\theta_q}^{\text{eff}}(\bm O;\bm\theta_q) \!=\! \bm C_q^{-1} \sum_{k=0}^K \bm\psi_{\bm\theta_q}^{(k)}(\bm O; \bm\theta_q)$, where $\bm C_q$ is a normalization matrix specified in Appendix C,\begingroup\makeatletter\def\f@size{10}\check@mathfonts 
\begin{align*}
\bm\psi_{\bm\theta_q}^{(K)}(\bm O;\bm\theta_q) & = \frac{d(\overline A_K,\bm Z;\bm\theta_q)}{\overline \pi_K(\overline A_K,\overline{\bm L}_K)}\left\{\mathbb{I}\left(Y\leq h(\overline A_K,\bm Z;\bm\theta_q)\right)-F_{Y_{\overline A_K}|\overline A_K,\overline{\bm L}_K}\left(h(\overline A_K,\bm Z;\bm\theta_q)\Big|\overline A_K,\overline{\bm L}_K\right)\right\},\\
\bm\psi_{\bm\theta_q}^{(k)}(\bm O;\bm\theta_q) & = \!\!\!\! \sum_{\underline{a}_{k+1} \!\in \underline{\mathbb{A}}_{k+1}} \!\!\! \frac{d(\overline A_k, \underline{a}_{k+1},\bm Z;\bm\theta_q)}{\overline \pi_k(\overline A_k,\overline{\bm L}_k)}\left\{F_{Y_{\overline A_{k+1},\underline{a}_{k+2}}|\overline A_{k+1},\overline{\bm L}_{k+1}}\left(h(\overline A_k, \underline{a}_{k+1},\bm Z;\bm\theta_q)\Big|\overline{A_k}, a_{k+1},\overline{\bm L}_{k+1}\right)\right.\\
& \quad \quad\quad \quad - F_{Y_{\overline A_{k},\underline{a}_{k+1}}|\overline A_{k},\overline{\bm L}_{k}}\left(h(\overline A_k, \underline{a}_{k+1},\bm Z;\bm\theta_q)\Big|\overline{A_k}, \overline{\bm L}_{k}\right)\Big\}, \quad \text{for }k=1,\dots,K-1, \\
\bm\psi_{\bm\theta_q}^{(0)}(\bm O;\bm\theta_q) & = \sum_{\overline a_K \in \overline{\mathbb A}_K} d(\overline a_K,\bm Z;\bm\theta_q) \left\{F_{Y_{A_1,\underline{a}_2}|A_1,\bm L_1}\left(h(\overline a_K,\bm Z;\bm\theta_q)\Big|a_1,\bm L_1\right)- q\right\}.
\end{align*}\endgroup
Therefore, the semiparametric efficiency bound for estimating $\bm\theta_q$ is $E\left[\mathbb{U}_{\bm\theta_q}^{\text{eff}}(\bm O;\bm\theta_{q0})^{\otimes 2}\right]$.
\end{theorem}

{\color{black}The derivation of Theorem \ref{EIF} follows \cite{bickel1993efficient} and details are given in Web Appendix C.} The EIF paves the way to construct a new estimator of $\bm\theta_q$ by solving $\mathbb{P}_n[\mathbb{U}_{\bm\theta_q}^{\text{eff}}(\bm O;\bm\theta_q)]=\bm 0$.
The EIF depends on two sets nuisance functions: propensity scores, $\pi_k(\overline A_K,\overline{\bm L}_k)$, $k=1,\dots,K$, and potential outcome distributions, $F_{Y_{\overline A_k,\underline{a}_{k+1}}|\overline A_k,\overline{\bm L}_k}$, $k=1,\dots,K$, as well as a constant normalization matrix $\bm C_q$ that does not affect the solution of $\mathbb{P}_n[\mathbb{U}_{\bm\theta_q}^{\text{eff}}(\bm O;\bm\theta_q)]=\bm 0$. This motivates us to find a new estimator by substituting the nuisance functions in the EIF, $h_{nuisance}=\big\{\pi_k(\overline A_k,\overline{\bm L}_k), F_{Y_{\overline A_k,\underline a_{k+1}}|\overline A_k,\overline{\bm L}_k},  \text{ for }k=1,\dots,K\big\}$, by their corresponding estimators in Section \ref{sec:ipw-icr}. Specifically, we can solve the EIF-based estimating equation:
\begin{equation}\label{ee:dr}
    \mathbb{P}_n\left[\mathbb{U}_{\bm \theta_q}^{\text{DR}}(\bm O; \bm\theta_q,\widehat{\bm \alpha},\widehat{\bm \beta})\right] = \bm 0,
\end{equation}
where $\mathbb{U}_{\bm \theta_q}^{\text{DR}}(\bm O;\bm\theta_q,\bm \alpha,\bm \beta)$ is given by
\begin{equation}\label{dr_estimating_score}
     \mathbb{U}_{\bm \theta_q}^{(K)} (\bm O; \bm\theta_q,\overline{\bm\alpha}_K,\bm\beta_K) + \sum_{k=1}^{K-1} \mathbb{U}_{\bm \theta_q}^{(k)} (\bm O; \bm\theta_q, \overline{\bm \alpha}_k, \bm \beta_{k},\bm \beta_{k+1}) + \mathbb{U}_{\bm \theta_q}^{(0)}(\bm O;\bm\theta_q,\bm \beta_1).
\end{equation}
Here, $\mathbb{U}_{\bm \theta_q}^{(k)}$, $k=0,\dots,K$, are $\bm\psi_{\bm\theta_q}^{(k)}$ in Theorem \ref{EIF} with $h_{nuisance}$ replaced by their parametric model representations $\{\pi_k(\overline A_k,\overline{\bm L}_k;\bm\alpha_k),\Psi_k(y,\overline A_k,\underline a_{k+1},\overline{\bm L}_k;\bm\beta_k),\text{ for } k=1,\dots,K\}$. In fact, as we will prove in Theorem \ref{thm:dr}, \eqref{ee:dr} is a doubly robust estimating equation in the sense that its expectation is $\bm 0$ when either $\mathcal M_{ps}$ or $\mathcal M_{om}$ is correctly specified, but not necessarily both.

The estimating equations used in the IPW and the doubly robust methods, i.e., \eqref{ee:ipw} and \eqref{ee:dr}, involve an indicator function $\mathbb{I}(Y \leq h(\overline a_K, \bm Z;\bm \theta_q))$, and therefore are discontinuous in $\bm\theta_q$. This may pose  challenges for solving the estimating equation and calculating the asymptotic variance. 
We propose a smoothed estimating equation approach to address the discontinuity issues, in a similar spirit to \cite{heller2007smoothed}. The smoothed estimating equation allows us to use a derivative-based algorithm to search for $\bm\theta_q$ and permit fast calculation of the standard errors via the sandwich method. 
Specifically, we use a local distribution function $\mathcal{K}\left((h(\overline A_K, \bm Z;\bm \theta_q)-Y)/\tau_n\right)$ to approximate the indicator function $\mathbb{I}(Y \leq h(\overline A_K, \bm Z;\bm \theta_q))$, where $\tau_n$ is a pre-specified bandwidth depending on the sample size $n$. The bandwidth $\tau_n$ converges to zero as $n$ increases such that if $Y < h(\overline A_K, \bm Z;\bm \theta_q)$, $\mathcal{K}\left((h(\overline A_K, \bm Z;\bm \theta_q)-Y)/\tau_n\right) \rightarrow 1$ as $n\rightarrow \infty$, and if $Y > h(\overline A_K, \bm Z;\bm \theta_q)$, $\mathcal{K}\left((h(\overline A_K, \bm Z;\bm \theta_q)-Y)/\tau_n\right) \rightarrow 0$ as $n\rightarrow \infty$. We require two regularity conditions for  $\mathcal{K}(x)$ and the bandwidth $\tau_n$: (B1) The local distribution function $\mathcal{K}(x)$ is continuous and its first-order derivative $\mathcal{K}'(x)=\frac{\partial}{\partial x} \mathcal{K}(x)$ is symmetric about 0 with $\int x^2 \mathcal{K}'(x) \text{d}x < \infty$; (B2) The bandwidth, $\tau_n$, is chosen such that as $n\rightarrow \infty$, $n\tau_n \rightarrow \infty$ and $n\tau_n^4 \rightarrow 0$. In this paper, we specify $\mathcal{K}(x)$ as the logistic function $\mathcal{K}(x)=e^x/(1+e^x)$. {\color{black}We set the bandwidth $\tau_n = \widehat{\sigma}n^{-0.26}$, where $\widehat{\sigma}$ is the sample standard deviation of the residual $Y-h(\overline A_K,\bm Z,\widehat{\bm\theta}_q)$, $\widehat{\bm\theta}_q$ is an initial guess of $\bm\theta_q$ (e.g, one can use $\widehat{\bm\theta}_q^{\text{ICR}}$), and the exponent $-0.26$ provides the quickest rate of convergence under the bandwidth constraint $n\tau_n^4 \rightarrow 0$ \citep{heller2007smoothed}. As is demonstrated in Theorem 2, this specification of the bandwidth does not affect the asymptotic behavior of smoothed estimators even if $\widehat{\bm\theta}_q^{\text{ICR}}$ is inconsistent, as long as $\widehat\sigma$ is positive, finite and satisfies (B2).}

To obtain the smoothed estimating equation, we replace the indicator function in the original estimating equation by the local distribution function. We solve the following smoothed doubly robust estimating equation with a derivative-based algorithm to obtain $\widehat{\bm\theta}_q^{\text{DR}}$:
\begin{align}
    & \mathbb{P}_n\left\{\widetilde{\mathbb{U}}_{\bm \theta_q}^{\text{DR}}(\bm O; \bm\theta_q,\widehat{\bm \alpha},\widehat{\bm \beta})\right\} =\bm 0 \label{see:dr}
\end{align}
where $\widetilde{\mathbb{U}}_{\bm \theta_q}^{\text{DR}}(\bm O; \bm\theta_q,\bm \alpha,\bm \beta)$ is \eqref{dr_estimating_score} with $\mathbb{U}_{\bm \theta_q}^{(K)}(\bm O; \bm\theta_q,\overline{\bm\alpha}_K, \bm\beta_K)$ replaced by \begingroup\makeatletter\def\f@size{10.5}\check@mathfonts
\begin{equation}\label{Tilde_U_K}
 \frac{d(\overline A_K,\bm Z;\bm\theta_q)}{\overline \pi_K(\overline A_K,\overline{\bm L}_K;\bm \alpha)}\left\{\mathcal{K}\left(\frac{ h(\overline A_K, \bm Z;\bm \theta_q)-Y}{\tau_n}\right)-\Psi_K\Big(h(\overline A_K,\bm Z;\bm\theta_q),\overline A_K, \overline{\bm L}_K;\bm\beta_K\Big)\right\}.
\end{equation}\endgroup
 Similarly, for the IPW, we can solve the following smoothed equation instead of \eqref{ee:ipw}:
\begin{equation}\label{see:ipw}
 \mathbb{P}_n \left[ \frac{d(\overline A_K,\bm Z;\bm\theta_q)}{ \overline{\pi}_K(\overline A_K, \overline{\bm L}_K,\widehat{\bm\alpha})} \left\{\mathcal{K}\left(\frac{ h(\overline A_K, \bm Z;\bm \theta_q)-Y}{\tau_n}\right)-q\right\}\right] = \bm 0,
\end{equation}
and we denote this new IPW estimator as $\widehat{\bm \theta}_q^{\text{IPW}}$. Theorem \ref{thm:dr} present the asymptotic properties of doubly robust estimators, obtained by the smoothed and the original unsmoothed estimating equation. For completeness, we also provide the asymptotic theory of the smoothed IPW estimator in Web Appendix D.



\begin{theorem}\label{thm:dr}
Suppose that either $\mathcal M_{ps}$ or $\mathcal M_{om}$ is correctly specified, and (A1)--(A4), (B1)--(B2), regularity conditions listed in Web Appendix E all hold. Then, (i) $\sqrt{n}(\widehat{\bm \theta}_q^{\text{DR}}-\widehat{\bm \theta}_q^{\text{DR}*})=o_p(1)$; (ii) $\sqrt{n}(\widehat{\bm \theta}_q^{\text{DR}} - \bm \theta_{q0})$ and $\sqrt{n}(\widehat{\bm \theta}_q^{\text{DR}*} - \bm \theta_{q0})$ converge in distribution to $N(\bm 0, \bm \Sigma^{\text{DR}})$, where $\bm \Sigma^{\text{DR}}$ is defined in Web Appendix E; (iii) if both $\mathcal M_{ps}$ and $\mathcal M_{om}$ are correctly specified, $\widehat{\bm \theta}_q^{\text{DR}}$ and $\widehat{\bm \theta}_q^{\text{DR}*}$ achieve the semiparametric efficiency bound and $\bm \Sigma^{\text{DR}}=E[\mathbb{U}_{\bm\theta_q}^{\text{eff}}(\bm O;\bm\theta_{q0})^{\otimes 2}]$. 
\end{theorem}

Proof of Theorem \ref{thm:dr} is given in Web Appendix E. It indicates that the doubly robust estimator obtained from the smoothed estimating equations are asymptotically equivalent to the original unsmoothed estimator; i.e., the smoothing process preserves the asymptotic efficiency and robustness of the original estimators, but simplifies the calculation. Also, Theorem \ref{thm:dr} formally states that the doubly robust estimators (both $\widehat{\bm \theta}_q^{\text{DR}}$ and $\widehat{\bm \theta}_q^{\text{DR}*}$) are semiparametric efficient when both $\mathcal M_{ps}$ and $\mathcal M_{om}$ are correctly specified and remain consistent when either $\mathcal M_{ps}$ or $\mathcal M_{om}$ is misspecified. Finally, another advantage of using the smoothed estimating equation is that it allows for fast computation of the asymptotic variance matrix, $\bm\Sigma^{\text{DR}}$, via the sandwich variance method. In Web Appendix E, we provide an explicit variance estimator for $\bm\Sigma^{\text{DR}}$. The finite sample performance of  $\widehat{\bm\theta}_q^{\text{DR}}$ will be investigated in Section~\ref{sec:simulation}. 

\begin{remark}
{\color{black}While our primary focus is on the parametric specification of the nuisance functions, it is feasible to enhance this estimator through double machine learning, which incorporates nonparametric modeling of nuisance functions \citep{chernozhukov2018double}. In Web Appendix F, we describe a double machine learning MSQM estimator with cross-fitting techniques (denoted by $\widehat{\bm\theta}_q^{\text{np}}$) and provide  results on the required rate conditions for consistency, asymptotic normality and efficiency. The results suggest that  $\widehat{\bm\theta}_{q}^{\text{np}}$ is CAN and efficient if all components in $h_{nusiance}$ are consistently estimated and the product of the convergence rate between the propensity score estimate and the outcome distribution estimate at the $k$-th time period, for all $k=1,\dots,K$, are at least $o_p(n^{-1/2})$ in $L_2(P)$-norm (Theorem S4 in Web Appendix F). In practice, the set of propensity scores $\{\pi_k(\overline A_k,\overline{\bm L}_k),k=1,\dots,K\}$ can be estimated using state-of-the-art machine learners for binary classification. However, machine learning estimation of the entire potential outcome distributions can be non-trivial and the theoretical properties of machine learning CDF estimators in general settings are relatively less studied. In Web Appendix F, we explain that, under a special case when all $\{\bm L_1,\dots,\bm L_K\}$ are categorical, it is feasible to pursue double machine learning by estimating $F_{Y_{\overline A_k,\underline a_{k+1}}|\overline A_k,\overline{\bm L}_k}$ with nonparametric estimators for conditional CDFs (e.g., Kernel smoothing, highly adaptive lasso \citep{hejazi2022haldensify}). However, if the time-varying covariates are high-dimensional with many continuous components, identifying a suitable machine learning CDF estimator for the potential outcome distributions requires future investigation.}
\end{remark}

\section{Sensitivity Analysis for Sequential Ignorability}\label{sec:sensitivity}

\subsection{Quantifying unmeasured confounding}

We propose a sensitivity analysis framework to assess the extent to which the violation of the no unmeasured confounding assumption (A2) might affect the estimation of $\bm\theta_q$. {\color{black}Following the general idea in \citet{brumback2004sensitivity} and \citet{hu2022flexible}, we proceed by first specifying a confounding function specific to MSQM to characterize the magnitude of unmeasured confounding and then developing bias-corrected estimators that specifically incorporate the confounding function.} For $k=1,\dots, K$, we define the confounding function
\begin{align}\label{sf_true}
c_k(y,\overline a_{K},\overline{\bm l}_k) & = F_{Y_{\overline a_K}|\overline A_k,\overline{\bm L}_k}(y|a_k,\overline a_{k-1},\overline{\bm l}_k) - F_{Y_{\overline a_K}|\overline A_{k},\overline{\bm L}_k}(y|1-a_k,\overline a_{k-1},\overline{\bm l}_{k}), 
\end{align}
which is a function of $y$ defined on the support of the potential outcome, $[y_{\min},y_{\max}]$. The confounding function measures, among the subgroup of population with $\overline A_{k-1}=\overline a_{k-1}$ and $\overline{\bm L}_k=\overline{\bm l}_k$, the difference in the potential outcome distribution between those receiving $A_k=a_k$ and those receiving $A_k=1-a_k$ in the $k$-th period. If there exists unmeasured confounding in the $k$-th period, then  $c_k(y,\overline a_{K},\overline{\bm l}_k) \neq 0$. The confounding function \eqref{sf_true} encodes the direction and magnitude of  unmeasured confounding for treatment allocation of $A_k$, conditional on the history of treatment and observed covariates at or before the beginning of the $k$-th period. An intuitive interpretation of \eqref{sf_true} with a single follow-up time ($K=1$) is provided in Table \ref{tab:interpretation}, which can be straightforwardly generalized to the settings with multiple time periods. 

\begin{table}[!p]
\centering
\scalebox{0.8}{\begin{threeparttable}
\caption{Interpretation of prior assumptions on the sign of $c_1(y,a_1,\bm l_1)$ with a fixed $y\in [y_{\min},y_{\max}]$, based on the context of our illustrative example in Section \ref{sec:examples}. Suppose a group of patients who developed severe hypertension were enrolled in the study, for whom a set of baseline confounders $\bm L_1$ were measured, and an antihypertensive treatment $A_1$ (1, treated; 0, untreated) was assigned at baseline. The outcome $Y$ is patient's systolic blood pressure (SBP, mmHg) at the end of period 1, where a higher $Y$ is undesirable. Since there is a single follow-up time ($K=1$), we have one confounding function $c_1(y, a_1,\bm l_1)$ quantifying the baseline unmeasured confounding. In this table, we provide interpretations of the confounding function based on the sign of $c_1(y, a_1,\bm l_1)$ with a fixed $y\in [y_{\min},y_{\max}]$. \label{tab:interpretation}}
\begin{tabular}{ccc}
\toprule
\multicolumn{2}{c}{Sign of $c_1(y, a_1,\bm l_1)$}  &  \multirow{2}{*}{Interpretation} \\
$c_1(y, 1,\bm l_1)$                & $c_1(y, 0,\bm l_1)$             &      \\ 
\midrule
$>0$ & $<0$ & \parbox{15cm}{The treated individuals have a higher probability to have their potential SBP below $y$ to both treatment and no treatment than untreated individuals, given the same level of measured covariates. In other words, healthier individuals were more likely to be treated, given the same level of measured covariates.} \\ \addlinespace
\midrule
$<0$ & $>0$ &   \parbox{15cm}{In contrary to the previous interpretation, unhealthier individuals were more likely to be treated, given the same level of measured covariates.} \\ \addlinespace
\midrule
$<0$ & $<0$ & \parbox{15cm}{The potential SBP to both treatment and no treatment has higher probability to be smaller than $y$ among individuals who choose it than among those who do not; that is, the observed treatment assignment is beneficial to the alternative which reverse treatment assignment.} \\ \addlinespace
\midrule
$>0$ & $>0$ & \parbox{15cm}{In contrary to the previous interpretation, the observed treatment assignment is undesirable to the alternative which reverse treatment assignment.} \\ \addlinespace
\bottomrule
\end{tabular}
\end{threeparttable}}
\end{table}

Typically, one can specify a working confounding function, $c_k(y,\overline a_K, \overline{\bm l}_k;\bm \gamma_k)$, indexed by a finite dimensional parameters $\bm \gamma_k$. 
To focus ideas, we require the following two regularity conditions for $c_k(y,\overline a_K, \overline{\bm l}_k;\bm \gamma_k)$. First, observing that the confounding function is a difference between two CDFs, we require that $c_k(y,\overline a_K, \overline{\bm l}_k;\bm\gamma_k)$ is continuous and bounded between $-1$ and $1$ for all $y \in [y_{\min},y_{\max}]$ and converges to 0 as $y\rightarrow y_{\min}$ or $y_{\max}$. Second, we assume the partial derivative of $c_k(y,\overline a_K, \overline{\bm l}_k;\bm\gamma_k)$ with respect to $y$, $c_k'(y,\overline a_K, \overline{\bm l}_k)=\frac{\partial}{\partial y} c_k(y,\overline a_K, \overline{\bm l}_k;\bm\gamma_k)$, exists and is continuous for $y \in [y_{\min},y_{\max}]$.  If the potential outcome $Y_{\overline a_K}$ is unbounded on $\mathbb{R}$, we propose a feasible working confounding function for \eqref{sf_true}, given by 
\begin{equation}\label{sf}
    c_k(y,\overline a_K, \overline{\bm l}_k;\bm\gamma_k) \!=\! r_k(\overline a_K, \overline{\bm l}_k;\bm\gamma_{k1})\exp\left\{-\pi\left[\frac{r_k(\overline a_K, \overline{\bm l}_k;\bm\gamma_{k1})\Big(y-m_k(\overline a_K, \overline{\bm l}_k;\bm\gamma_{k3})\Big)}{b_k(\overline a_K, \overline{\bm l}_k;\bm\gamma_{k2})}\right]^2\right\},
\end{equation}
where $\pi$ is the mathematical constant, $\bm\gamma_k=[\bm\gamma_{k1}^T,\bm\gamma_{k2}^T,\bm\gamma_{k3}^T]^T$, and $r_k(\overline a_K, \overline{\bm l}_k;\bm\gamma_{k1})$, $b_k(\overline a_K, \overline{\bm l}_k;\bm\gamma_{k2})>0$, and $m_k(\overline a_K, \overline{\bm l}_k;\bm\gamma_{k3})$ are the \textit{peak}, \textit{area} and \textit{location} components, respectively. In Appendix G, we show that the working confounding function \eqref{sf} is a good approximation to the true confounding function \eqref{sf_true} when $Y_{\overline a_K}$ given $\overline A_k$ and $\overline{\bm L}_k$ is normally distributed and the two distributions used in the confounding function only have a mean shift.

\newcommand*{\num}{pi}
\usetikzlibrary{decorations.pathreplacing}
\tikzset{elegant/.style={smooth,thick,samples=50,cyan}}
\tikzset{eaxis/.style={->,>=stealth}}

\captionsetup[subfigure]{singlelinecheck=false}

\begin{figure}[h]
\centering
\begin{subfigure}[b]{1\textwidth}
\caption[Ex1]{Graph of $c_k(y,\overline a_K, \overline{\bm l}_k;\bm\gamma_k)$ when $r_k(\overline a_K, \overline{\bm l}_k;\bm\gamma_{k1})>0$.}
\centering
\begin{tikzpicture}[scale=2.5]
    \draw[eaxis] (-1.3,0) -- (\num,0) node[below] {{\Large $y$}};
   \draw [gray, line width=0.5mm, dotted] (-1,1.3) -- (\num,1.3);
    \node[text width=0.7cm] at (-1,1.3) {{\scriptsize 1}};
    \draw[eaxis] (-1,-0.2) -- (-1,1.5); 
    \node[text width=6.6cm] at (-1,1.5) {{\large $c_k(y,\overline a_K, \overline{\bm l}_k;\bm\gamma_k)$}};
    \draw[elegant,black,fill=gray!20,domain=-1.2:(\num-0.1)] plot(\x, {exp(-(\x-1)*(\x-1))});
    \draw [gray, line width=0.5mm, loosely dotted] (1,1) -- (1,0) node[below]{{\color{black}{\scriptsize $m_k(\overline a_K, \overline{\bm l}_k;\bm\gamma_{k3})$}}};
    \draw [gray, line width=0.5mm,loosely dotted] (1,1) -- (-1,1) node[left]{ {\color{black}{\scriptsize $r_k(\overline a_K, \overline{\bm l}_k;\bm\gamma_{k1})$}}};
    \draw[->, gray, line width=0.3mm] (1.5,0.5) -- (1.8,0.7) node[right]{{\color{black}{\scriptsize shaded area =$b_k(\overline a_K, \overline{\bm l}_k;\bm\gamma_{k2})$}}};
\end{tikzpicture}    
\end{subfigure}
\begin{subfigure}[b]{1\textwidth} 
\vspace{0.5cm}
\centering
\caption[Ex1]{Graph of $c_k(y,\overline a_K, \overline{\bm l}_k;\bm\gamma_k)$ when $r_k(\overline a_K, \overline{\bm l}_k;\bm\gamma_{k1})<0$.}
\begin{tikzpicture}[scale=2.5]
    \draw[eaxis] (-1.3,0) -- (\num,0) node[below] {{\Large $y$}};
    \draw [gray, line width=0.5mm, dotted] (-1,-1.3) -- (\num,-1.3);
    \node[text width=0.7cm] at (-1,-1.3) {{\scriptsize -1}};
    \draw[eaxis] (-1,-1.5) -- (-1,0.3); 
    \node[text width=6.6cm] at (-1,0.3) {{\large $c_k(y,\overline a_K, \overline{\bm l}_k;\bm\gamma_k)$}};
    \draw[elegant,black,fill=gray!20,domain=-1.2:(\num-0.1)] plot(\x, {-exp(-(\x-1)*(\x-1))});
    \draw [gray, line width=0.5mm, loosely dotted] (1,-1) -- (1,0) node[above]{{\color{black}{\scriptsize $m_k(\overline a_K, \overline{\bm l}_k;\bm\gamma_{k3})$}}};
    \draw [gray, line width=0.5mm,loosely dotted] (1,-1) -- (-1,-1) node[left]{ {\color{black}{\scriptsize $r_k(\overline a_K, \overline{\bm l}_k;\bm\gamma_{k1})$}}};
    \draw[->, gray, line width=0.3mm] (1.5,-0.5) -- (1.8,-0.7) node[right]{{\color{black}{\scriptsize shaded area =$b_k(\overline a_K, \overline{\bm l}_k;\bm\gamma_{k2})$}}};
\end{tikzpicture}    
\end{subfigure}
\caption[ Ex ]
{\small Graphical illustration of the sensitivity function $c_k(y,\overline a_K, \overline{\bm l}_k;\bm\gamma_k)$, when the \textit{peak} parameter $r_k(\overline a_K, \overline{\bm l}_k;\bm\gamma_{k1})>0$ (Panel a) or $<0$ (Panel b).} 
\label{fig:sf}
\end{figure}
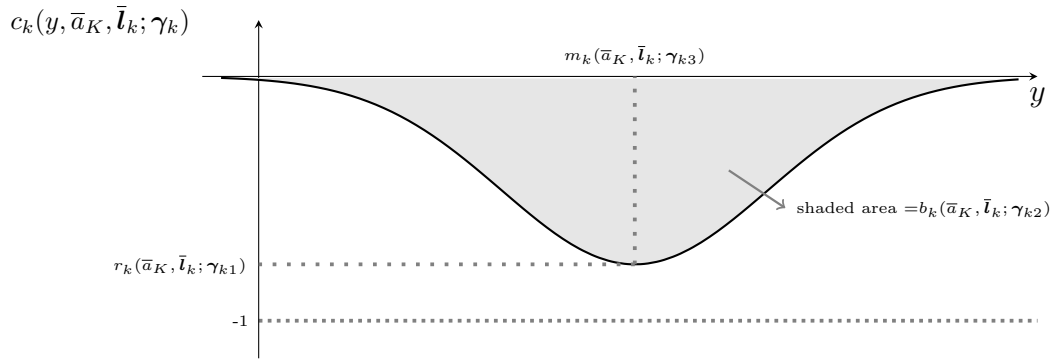

As shown in Figure \ref{fig:sf}, the working confounding function defined in \eqref{sf} is a bell or inverted bell curve that reaches the extremum value, $r_k(\overline a_K, \overline{\bm l}_k;\bm\gamma_{k1})$, at the \textit{location}, $m_k(\overline a_K, \overline{\bm l}_k;\bm\gamma_{k3})$, and then decreases to zero when $y \rightarrow \pm \infty$. Therefore, the \textit{peak} component, $r_k(\overline a_K, \overline{\bm l}_k;\bm\gamma_{k1})$, measures the maximum difference between the two CDFs used in the confounding function \eqref{sf_true}, $F_{Y_{\overline a_K}|\overline A_k,\overline{\bm L}_k}(y|\overline a_k,\overline{\bm l}_k)$ and $F_{Y_{\overline a_K}|\overline A_{k},\overline{\bm L}_k}(y|1-a_k,\overline a_{k-1},\overline{\bm l}_{k})$ (abbreviated by $F_k^{(1)}(y)$ and $F_k^{(0)}(y)$ below). The absolute value of $r_k(\overline a_K, \overline{\bm l}_k;\bm\gamma_{k1})$ is $\underset{y}{\text{sup}}|c_k(y,\overline a_K, \overline{\bm l}_k;\bm\gamma_k)|=\underset{y}{\text{sup}}|F_k^{(1)}(y)-F_k^{(0)}(y)|$, which is the total variation distance between $F_k^{(1)}$ and $F_k^{(0)}$. Notice that $r_k(\overline a_K, \overline{\bm l}_k;\bm\gamma_{k1})$ should be bounded between $-1$ and $1$ since it is a difference between two probabilities. In particular, $r_k(\overline a_K, \overline{\bm l}_k;\bm\gamma_{k1})=0$ suggests $c_k(y,\overline a_K, \overline{\bm l}_k;\bm\gamma_k) \equiv 0$ for $y \in \mathbb{R}$, i.e., no unmeasured confounding. If the \textit{peak} component is not zero, the \textit{area} component, $b_k(\overline a_K, \overline{\bm l}_k;\bm\gamma_{k2})$, measures the area under the confounding function curve $c_k(y,\overline a_K, \overline{\bm l}_k;\bm\gamma_k)$; that is, $b_k(\overline a_K, \overline{\bm l}_k;\bm\gamma_{k2})=\int_{y} |c_k(y,\overline a_K, \overline{\bm l}_k;\bm\gamma_k)| dy=\int_{y} |F_k^{(1)}(y)-F_k^{(0)}(y)| dy$, which is the $1-$Wasserstein distance (\citealp{de20211}) between the two CDFs. The \textit{peak} and \textit{area} components quantify the distance between $F_k^{(1)}$ and $F_k^{(0)}$, but from complementary perspectives. The \textit{peak} searches the maximum difference between $F_k^{(1)}(y)$ and $F_k^{(0)}(y)$ among all possible $y\in \mathbb R$, whereas the \textit{area} integrates the absolute difference between $F_k^{(1)}(y)$ and $F_k^{(0)}(y)$ on $y\in \mathbb R$. 

\subsection{Bias-corrected estimators under unmeasured confounding}

Leveraging $c_k(y,\overline a_k,\overline{\bm l}_k;\bm\gamma_k)$, $k=1,\dots,K$, we can generalize the IPW, ICR, and doubly robust approaches to bias-corrected estimators to enable consistent estimation of $\bm\theta_q$ after accounting for unmeasured confounding. {\color{black}To conduct sensitivity analysis, one can choose $\bm\gamma = [\bm\gamma_1^T,\dots,\bm\gamma_K^T]^T$ from a predefined  set of values $\bm\Gamma$ and report the point estimates and confidence intervals for each $\bm\gamma \in \bm\Gamma$, which summarizes the impact due to violations of assumption (A2). We note that the number of sensitivity parameters grows as the number of periods increases, which may complicate both the implementation and  interpretation of the analysis. When $K$ is small, we may separately consider sensitivity analysis for each $k$ to disentangle the impact due to period-specific violation of ignorability (see Section \ref{sec:examples} for an example). When $K$ is large, it could be helpful to simplify the sensitivity analysis by reducing the number of sensitivity parameters. For example, assuming that the amount of unmeasured confounding is similar across $K$ periods, we can set $\bm\gamma_1=\bm\gamma_2=\cdots=\bm\gamma_K$ such that only one set of sensitivity parameters is required. Below, we pursue a general methodology and present} the bias-corrected doubly robust estimator ($\widehat{\bm \theta}_q^{\text{BC-DR}}$); bias-corrected IPW and ICR estimators ($\widehat{\bm \theta}_q^{\text{BC-IPW}}$ and $\widehat{\bm \theta}_q^{\text{BC-ICR}}$) can be similarly constructed and provided in Web Appendix H.

Defining $c_k^*(y,\overline a_K, \overline{\bm l}_k;\bm\gamma_k,\bm\alpha_k)=c_k(y,\overline a_K, \overline{\bm l}_k;\bm\gamma_k) \times \pi_k(1-a_k,\overline{a}_{k-1},\overline{\bm l}_k;\bm \alpha_k)$, the bias-corrected doubly robust estimator $\widehat{\bm \theta}_q^{\text{BC-DR}}$ is given by the solution of the smoothed estimating equation:
\begingroup\makeatletter\def\f@size{10}\check@mathfonts\begin{align*}
     \mathbb{P}_n\left\{\widetilde{\mathbb{U}}_{\bm \theta_q}^{(K)} (\bm O; \bm\theta_q,\widehat{\overline{\bm\alpha}}_K,\widehat{\bm\beta}_K) + \sum_{k=1}^{K-1} \widetilde{\mathbb{U}}_{\bm \theta_q}^{(k)} (\bm O; \bm\theta_q,\widehat{\overline{\bm \alpha}}_{k+1}, \widehat{\bm \beta}_{k},\widehat{\bm \beta}_{k+1}) + \widetilde{\mathbb{U}}_{\bm \theta_q}^{(0)}(\bm O; \bm\theta_q,\widehat{\bm \alpha}_1,\widehat{\bm \beta}_1)\right\} = \bm 0, \label{see:adr}
\end{align*}\endgroup
where $\widehat{\bm\alpha}_k$'s are obtained in Section \ref{sec:IPW-alone}, $\widehat{\bm \beta}_k$'s are obtained from the bias-corrected ICR approach (Web Appendix H), $\widetilde{\mathbb{U}}_{\bm \theta_q}^{(K)} (\bm O; \bm\theta_q,\overline{\bm\alpha}_K,\bm\beta_K)$ is given in \eqref{Tilde_U_K}, and
\begingroup\makeatletter\def\f@size{9.5}\check@mathfonts
\begin{align*}
    \widetilde{\mathbb{U}}_{\bm \theta_q}^{(k)} (\bm O; \bm\theta_q,\!\overline{\bm \alpha}_{k\!+\!1}, \bm \beta_{k},\!\bm \beta_{k+1})\! = &  \sum_{\underline{a}_{k\!+\!1} \in \underline{\mathbb{A}}_{k\!+\!1}} \!\!\! \frac{d(\overline A_k, \underline{a}_{k+1},\bm Z;\bm\theta_q)}{\overline \pi_k(\overline A_k,\overline{\bm L}_k;\overline{\bm \alpha}_k)} \left\{\Psi_{k+1}\Big(h(\overline A_k, \underline{a}_{k+1},\bm Z;\bm\theta_q),\overline A_k, \underline{a}_{k+1}, \overline{\bm L}_{k+1};\bm\beta_{k+1}\Big)\right.  \\
    & \quad \quad  -c_{k+1}^*\Big(h(\overline A_k, \underline{a}_{k+1},\bm Z;\bm\theta_q),\overline A_k,\underline a_{k+1},\overline{\bm L}_{k+1};\bm\gamma_{k+1},\bm\alpha_{k+1}\Big)\\  & -  \left. \Psi_k\Big(h(\overline A_k, \underline a_{k+1},\bm Z;\bm \theta_q),\overline A_k, \underline a_{k+1},\overline{\bm L}_{k};\bm\beta_{k}\Big) \right\}\text{ for } k=1,\dots,K-1,
\end{align*}
$$
\widetilde{\mathbb U}_{\bm \theta_q}^{(0)} (\bm O; \bm\theta_q,\bm\alpha_1,\bm \beta_1) \!=  \!\!\!\! \sum_{\overline a_K \in \overline{\mathbb A}_K} \!\!d(\overline a_K,\bm Z;\bm\theta_q) \left\{\Psi_1\Big(h(\overline a_K,\bm Z;\theta_q),\overline a_K,\bm L_1;\bm\beta_1\Big)\!-\!c_{1}\Big(h(\overline a_K,\bm Z;\bm\theta_q),\overline a_K,\bm L_{1};\bm\gamma_1,\bm\alpha_1\Big) \!-\! q\right\}.
$$
\endgroup
Compared to estimating equation \eqref{see:dr}, the bias-corrected estimating equation includes extra terms, $c_k^*$, to remove the hidden bias due to unmeasured confounding. 

We develop the the asymptotic theory of $\widehat{\bm \theta}_q^{\text{BC-IPW}}$, $\widehat{\bm \theta}_q^{\text{BC-ICR}}$, and $\widehat{\bm \theta}_q^{\text{BC-DR}}$ and summarize their properties in Theorem S3 of Web Appendix H. In particular, $\widehat{\bm\theta}_q^{\text{BC-ICR}}$ requires that both $\mathcal M_{ps}$ and $\mathcal M_{om}$ are correctly specified, whereas $\widehat{\bm\theta}_q^{\text{BC-IPW}}$ and $\widehat{\bm\theta}_q^{\text{BC-DR}}$ only require $\mathcal M_{ps}$ is correctly specified. This suggests that $\widehat{\bm\theta}_q^{\text{BC-DR}}$ is consistent if the propensity score models based only on the observed covariates (i.e., marginalized over the unmeasured confounders) are correctly specified, regardless of whether the outcome regression models are correct or not. However, we may expect that $\widehat{\bm\theta}_q^{\text{BC-DR}}$ is still more efficient than the $\widehat{\bm\theta}_q^{\text{BC-IPW}}$ when the outcome models are not substantially misspecified. Consistent estimators of the asymptotic variances of $\widehat{\bm \theta}_q^{\text{BC-IPW}}$, $\widehat{\bm \theta}_q^{\text{BC-ICR}}$, and $\widehat{\bm \theta}_q^{\text{BC-DR}}$ are given in Web Appendix H.

\section{Simulation Studies}\label{sec:simulation}

\subsection{Simulations without unmeasured confounders}\label{sec:simulation_sub1}

We conducted simulations to evaluate the finite-sample performance of the $\widehat{\bm\theta}_q^{\text{IPW}}$, $\widehat{\bm\theta}_q^{\text{ICR}}$ and $\widehat{\bm\theta}_q^{\text{DR}}$, when (A2) holds. We considered $K=3$ time periods with a binary treatment $A_k \in \{0,1\}$ assigned at period $k$. The observed data $\bm O = \Big\{\bm L_1 =(L_{11},L_{12}),A_1,\bm L_2 = (L_{21},L_{22}),A_2,\bm L_3 = (L_{31},L_{32}),A_3,Y\Big\}$ were simulated based on the data generation process in Web Appendix I. We considered two scenarios with different levels of overlap (Web Figure S1). In Scenario I, the true propensity score distributions at each time period have moderate overlap between comparison groups; in Scenario II, the true propensity score distributions have weak overlap. Under (A1)--(A4), we show in Web Appendix I that 
the marginal distribution of $Y_{\overline a_3}$ follows $N(10-4a_1-4a_2-10a_3,28+12a_3)$. We are interested in estimating the MSQM $Q_{Y_{\overline a_3}}^{(q)} = \theta_{q,0}+\theta_{q,1}a_1+\theta_{q,2}a_2+\theta_{q,3}a_3$ with unknown parameter $\bm\theta_q=[\theta_{q,0},\theta_{q,1},\theta_{q,2},\theta_{q,3}]$. The true value for $\bm\theta_q$ is $\theta_{q,0} = 10+5.292z_q$, $\theta_{q,1}=\theta_{q,2}=-4$, and $\theta_{q,3}=-10+1.033z_q$, where $z_q$ is the $q$-th lower-quantile of the standard normal distribution. 



We conducted 1,000 replications, each with a sample size of 2,000, to evaluate the percent bias, Monte Carlo standard error, and the 95\% confidence interval coverage of each causal estimator. 
In addition, we employed two quantile regression (QR) estimators, $\widehat{\bm\theta}_q^{U}$ and $\widehat{\bm\theta}_q^{A}$, to benchmark the bias resulting from the failure to properly account for confounding by $\overline{\bm L}_3$. The unadjusted estimator ($\widehat{\bm\theta}_q^{U}$) is based on QR with only the time-varying treatments $(A_1,A_2,A_3)$ as linear predictors. In contrast, the adjusted estimator ($\widehat{\bm\theta}_q^{A}$) implements QR with all time-varying treatments and confounders as linear predictors. We compared two methods for calculating the 95\% confidence interval, one based on a Wald-type approach with the asymptotic variance estimator and the other based on a non-parametric percentile bootstrap. For each causal estimator, specifications of the correct and incorrect nuisance models are given in Web Table S1. We evaluated the performance of $\widehat{\bm \theta}_q$ under $q\in\{0.25,0.5,0.75\}$.

\begin{table}[ht]
\scalebox{0.8}{\begin{threeparttable}
\caption{Simulation results for estimating $\bm\theta_{0.5}=[\theta_{0.5,0},\theta_{0.5,1},\theta_{0.5,2},\theta_{0.5,3}]$ in Scenario I (the treatment-specific propensity scores are moderately overlapped). We considered both correct (denoted by `T') or incorrect (denoted by `F') specifications of the propensity score models ($\mathcal M_{ps}$) and the outcome regression models ($\mathcal M_{om}$). \label{tab:sim_main} }
\centering
\begin{tabular}{llllrrrrrrrr}
  \toprule
 & & \multicolumn{2}{c}{Specification} & \multicolumn{4}{c}{Percent bias (\%)} & \multicolumn{4}{c}{Standard error} \\
  \cmidrule(r){3-4} \cmidrule(r){5-8} \cmidrule(r){9-12}
Model & Method & $\mathcal M_{ps}$ & $\mathcal M_{om}$ & $\widehat\theta_{0.5,0}$ & $\widehat\theta_{0.5,1}$ & $\widehat\theta_{0.5,2}$ & $\widehat\theta_{0.5,3}$ & $\widehat\theta_{0.5,0}$ & $\widehat\theta_{0.5,1}$ & $\widehat\theta_{0.5,2}$ & $\widehat\theta_{0.5,3}$ \\ 
  \hline
QR & Unadjusted &  &  & $-$34.16 & $-$61.92 & $-$52.39 & $-$22.46 & 0.27 & 0.31 & 0.30 & 0.30 \\ 
&  Adjusted &  &  & $-$0.05 & 149.95 & 149.83 & $-$0.07 & 0.19 & 0.36 & 0.35 & 0.27 \\
MSQM & IPW & T &  & $-$0.22 & $-$1.07 & 0.21 & $-$0.27 & 0.60 & 0.61 & 0.60 & 0.63 \\ 
 & IPW & F &  & $-$35.97 & $-$67.92 & $-$54.13 & $-$20.90 & 0.24 & 0.28 & 0.28 & 0.29 \\ 
 & ICR &  & T & $-$0.15 & $-$0.11 & $-$0.26 & $-$0.06 & 0.21 & 0.26 & 0.21 & 0.15 \\ 
 & ICR &  & F & $-$14.44 & $-$34.18 & $-$18.19 & $-$8.00 & 0.21 & 0.26 & 0.24 & 0.21 \\ 
 & DR & T & T & 0.08 & 0.78 & $-$0.26 & 0.04 & 0.35 & 0.45 & 0.43 & 0.39 \\ 
 & DR & T & F & $-$0.22 & 0.29 & $-$0.10 & 0.15 & 0.45 & 0.53 & 0.49 & 0.51 \\ 
 & DR & F & T & 1.32 & 2.38 & 1.51 & 0.69 & 0.23 & 0.29 & 0.25 & 0.20 \\ 
 & DR & F & F & $-$15.52 & $-$36.04 & $-$20.27 & $-$7.17 & 0.26 & 0.29 & 0.27 & 0.25 \\ 
  \hline
 & & \multicolumn{2}{c}{Specification} & \multicolumn{4}{c}{Coverage rate$^1$ (\%)} & \multicolumn{4}{c}{Coverage rate$^2$ (\%)} \\
  \cmidrule(r){3-4} \cmidrule(r){5-8} \cmidrule(r){9-12} 
Model & Method & $\mathcal M_{ps}$ & $\mathcal M_{om}$ & $\widehat\theta_{0.5,0}$ & $\widehat\theta_{0.5,1}$ & $\widehat\theta_{0.5,2}$ & $\widehat\theta_{0.5,3}$ & $\widehat\theta_{0.5,0}$ & $\widehat\theta_{0.5,1}$ & $\widehat\theta_{0.5,2}$ & $\widehat\theta_{0.5,3}$  \\ 
  \cmidrule(r){1-12}
 QR & Unadjusted &  &  & 0.0 & 0.0 & 0.0 & 0.0 & 0.0 & 0.0 & 0.4 & 0.2 \\ 
 & Adjusted &  &  & 97.5 & 0.0 & 0.0 & 96.5 & 97.3 & 0.0 & 0.0 & 97.5 \\ 
 MSQM & IPW & T &  & 93.0 & 94.0 & 95.0 & 93.4 & 93.8 & 94.7 & 94.2 & 95.1 \\ 
 & IPW & F &  & 0.0 & 0.0 & 0.0 & 0.0 & 0.0 & 0.0 & 0.2 & 0.3 \\ 
 & ICR &  & T & 94.9 & 94.3 & 95.6 & 95.5 & 94.4 & 94.8 & 95.4 & 95.9 \\ 
 & ICR &  & F & 0.0 & 0.1 & 13.2 & 3.1 & 1.2 & 6.3 & 47.5 & 30.2 \\ 
 & DR & T & T & 93.7 & 94.2 & 94.1 & 94.7 & 94.8 & 96.5 & 94.7 & 95.2 \\ 
 & DR & T & F & 96.3 & 94.2 & 95.5 & 94.8 & 96.1 & 96.7 & 95.0 & 95.6 \\ 
 & DR & F & T & 92.8 & 95.0 & 94.5 & 94.3 & 92.7 & 94.8 & 94.2 & 94.8 \\ 
 & DR & F & F & 0.0 & 0.1 & 15.8 & 19.9 & 2.9 & 11.5 & 57.5 & 56.3 \\ 
    \bottomrule
\end{tabular}
\begin{tablenotes}
    \item[*] The percent bias was calculated as the mean of the ratio of bias to the true value over 1,000 replications, i.e., $\text{Bias(\%)}=mean(\frac{\widehat{p}-p}{p}) \times 100\%$, where $p$ denotes the true value of the causal parameter and $\widehat{p}$ is its point estimate. The standard error is defined as the squared root of the empirical variance of causal parameter from the 1,000 replications. The coverage rate$^1$ is by a Wald-type 95\% confidence interval using the derived asymptotic variance formula. The coverage rate$^2$ is by a non-parametric percentile bootstrapping confidence interval using the 2.5\% and 97.5\% percentiles of the bootstrap distribution.  
  \end{tablenotes}
\end{threeparttable}}
\end{table}

Simulation results of $\widehat{\bm\theta}_{0.5}$ under scenario I (moderate overlap) are summarized in Table \ref{tab:sim_main}. Both QR estimators provide undesirable results due to failure in properly accounting for time-dependent confounding. The unadjusted QR estimator ($\widehat{\bm\theta}_{0.5}^{U}$) always exhibits notable bias with attenuated coverage. Although the adjusted QR estimator ($\widehat{\bm\theta}_{0.5}^{A}$) happens to carry small bias for $\theta_{0.5,0}$ and $\theta_{0.5,3}$, its bias for $\theta_{0.5,1}$ and $\theta_{0.5,2}$ is still substantial. In general, $\widehat{\bm \theta}_{0.5}^{\text{IPW}}$ yielded minimal bias with nominal coverage rates when $\mathcal M_{ps}$ was correctly specified but was considerably biased when $\mathcal M_{ps}$ was misspecified. Similarly, $\widehat{\bm \theta}_{0.5}^{\text{ICR}}$ was nearly unbiased and had the smallest variance when $\mathcal M_{om}$ were correctly specified but had a substantial bias otherwise. Furthermore, $\widehat{\bm \theta}_{0.5}^{\text{DR}}$, carried minimal bias if either $\mathcal M_{om}$ or $\mathcal M_{ps}$ was correct; The confidence interval based on our asymptotic variance estimator had nominal coverage rate when either $\mathcal M_{om}$ or $\mathcal M_{ps}$ were correct. When all nuisance models were correctly specified, $\widehat{\bm \theta}_{0.5}^{\text{ICR}}$ had the smallest Monte Carlo standard error, followed by $\widehat{\bm \theta}_{0.5}^{\text{DR}}$, whereas $\widehat{\bm \theta}_{0.5}^{\text{IPW}}$ is the least efficient. Interestingly, the Monte Carlo standard error of $\widehat{\bm \theta}_{0.5}^{\text{DR}}$ was consistently smaller than that of $\widehat{\bm \theta}_{0.5}^{\text{IPW}}$, even if either $\mathcal M_{om}$ or $\mathcal M_{ps}$ was misspecified. Simulation results for $\bm \theta_{0.75}$ and $\bm \theta_{0.25}$ are qualitatively similar and provided in Web Tables S2--S3.

Under weak overlap (Scenario II), the presence of extreme propensity scores inflated the finite-sample bias the IPW and doubly robust estimators (Web Tables S4--S6). However, the bias and Monte Carlo standard error of $\widehat{\bm\theta}_q^{\text{DR}}$ remained smaller than those of $\widehat{\bm\theta}_q^{\text{IPW}}$, suggesting that the former is more robust against weak overlap than the latter.

\subsection{Simulations in the presence of unmeasured confounding}\label{sec:simulation_sub2}

We carried out simulations to assess the effectiveness of the bias-corrected estimators, $\widehat{\bm\theta}_q^{\text{BC-IPW}}$, $\widehat{\bm\theta}_q^{\text{BC-ICR}}$, and $\widehat{\bm\theta}_q^{\text{BC-DR}}$, in reducing bias due to unmeasured confounding. The data generation process mirrors Scenario I in Section \ref{sec:simulation_sub1}, except that certain time-varying covariates are now considered unobserved. To isolate the impact of unmeasured confounding at different time periods, we examined three distinct cases: (I) Case 1: unmeasured baseline confounding, where one baseline covariate, $L_{11}$, is unobserved but all other covariates are measured; (II) Unmeasured confounding at the second period, where $L_{21}$ is unobserved, but all other covariates are measured; (III) Case 3: Unmeasured confounding at the third period, where $L_{31}$ is unobserved, but all other covariates remain available. We evaluated the performance of all three bias-corrected MSQM estimators with $q=0.5$, and compared the results with those from the uncorrected estimators. The specification of $\mathcal M_{ps}$ and $\mathcal M_{om}$ follows Section \ref{sec:simulation_sub1}, except that we eliminated the unmeasured variable from the working models. For instance, in Case 3, we removed all terms involving $L_{31}$ from all logistic models for $\mathcal M_{ps}$ and all linear models for $\mathcal M_{om}$. We used a sufficient large simulated data set with 2 million observations, along with the working confounding function \eqref{sf} to approximate true confounding functions \eqref{sf_true} in each Case; the details are provided in Web Appendix I.

The simulation results for Case 1 are presented in Web Table S7. The uncorrected IPW, ICR, and doubly robust estimators present non-negligible bias with attenuated coverage for estimating $\theta_{0.5,0}$ and $\theta_{0.5,1}$, even when both $\mathcal M_{ps}$ and $\mathcal M_{om}$ are both correctly specified. These uncorrected estimators had relatively small bias for estimating $\theta_{0.5,2}$ and $\theta_{0.5,3}$. This is because the unmeasured confounder in Case 1 only appears at the first time period and primarily influences the estimation of the structural coefficient for $A_1$ ($\theta_{0.5,1}$). The performance of the bias-corrected estimators, $\widehat{\bm\theta}_{0.5}^{\text{BC-IPW}}$, $\widehat{\bm\theta}_{0.5}^{\text{BC-ICR}}$, and $\widehat{\bm\theta}_{0.5}^{\text{BC-DR}}$, aligns with the their theoretical properties in Theorem S3 in Web Appendix H. Specifically, $\widehat{\bm\theta}_{0.5}^{\text{BC-IPW}}$ and $\widehat{\bm\theta}_{0.5}^{\text{BC-ICR}}$ exhibit small bias and nominal coverage when $\mathcal M_{ps}$ and $\mathcal M_{om}$ are correct, respectively. Furthermore, $\widehat{\bm\theta}_{0.5}^{\text{BC-DR}}$ demonstrates small bias with nominal coverage when either $\mathcal M_{ps}$ or $\mathcal M_{om}$ is correctly specified. Across the scenarios, $\widehat{\bm\theta}_{0.5}^{\text{BC-ICR}}$ and $\widehat{\bm\theta}_{0.5}^{\text{BC-DR}}$  appear  robust against misspecification of $\mathcal M_{ps}$, although this may not hold under other data generating processes. Simulation results for Cases 2--3 are generally similar and found in Web Tables S8--S9. 

\section{Analyses of Yale New Haven Health System Electronic Health Record data}\label{sec:examples}

\subsection{Estimation of quantile causal effects} 

We applied the proposed methods to the Yale New Haven Health System (YNHHS) Electronic Health Record data, a cohort study investigating the blood pressure (BP) response to prescribed antihypertensive for inpatients who  developed severe inpatient hypertension (systolic BP (SBP) $>$ 180 mmHg or diastolic BP (DBP) $>$ 110 mmHg) after admission (\citealp{ghazi2022severe}). The study includes adult patients admitted to one of the five YNHHS hospitals between 2016 to 2020. The outcome is the BP responses over 6 hours following the development of severe hypertension (HTN). We estimate the effect of antihypertensive treatments given at $(0,2)$, $[2,4)$, and $[4,6)$ hours after the patients developing severe HTN on quantiles of the patients' SBP at 6 hour following development of severe HTN.

This analysis included $K=3$ time periods, $(0,2)$, $[2,4)$, and $[4,6)$ hours following development of severe HTN, corresponding to $k=1,2$ and 3. At each time period, we code the antihypertensive treatment $A_k$ as 1 if an antihypertensive was given during this time period and 0 otherwise. The baseline covariates, $\bm L_1$, consist of age, gender, race, body mass index, comorbidities including history on cardiovascular and hypertension diseases, and SBP and DBP measurements at the time of developing severe HTN. The time-varying covariates, $\bm L_2$ (and $\bm L_3$), include the the most recent measurements on the patient's SBP and DBP prior to 2 (and 4) hours following development of severe HTN. The outcome $Y$ is the SBP measured at 6 hours after developing severe HTN. A total of 5,925 individuals who did not have missing data on any variables are included into this analysis. The MSQM considered here is $Q_{Y_{\overline a_3}}^{(q)} = \theta_{q,0} + \sum_{k=1}^3\theta_{q,k}a_k$ with causal parameters $\bm \theta_q = [\theta_{q,0},\theta_{q,1},\theta_{q,2},\theta_{q,3}]^T$. We employ logistic regressions for the propensity scores, adjusting for the main effects of the time-varying treatment and covariates $\overline A_{k-1}$ and $\overline{\bm L}_k$, the quadratic term of the continuous components of $\overline{\bm L}_k$, and all pairwise interaction terms between $\overline A_{k-1}$ and $\overline{\bm L}_k$. We utilize heteroscedastic linear regressions to model the potential outcomes. For the conditional mean, $\bm g_{k}(\overline A_k,\underline{a}_{k+1},\overline{\bm L}_k)$, we include the main effects of $\{\overline A_k,\underline{a}_{k+1},\overline{\bm L}_k\}$, the quadratic term for continuous components in $\overline{\bm L}_k$, and all pairwise interactions between $\{\overline A_k,\underline{a}_{k+1}\}$ and $\overline{\bm L}_k$. Regarding the conditional variance, $\sigma_k^2(\overline A_k,\underline{a}_{k+1},\overline{\bm L}_k;\bm\eta_k)$, we apply a log-transformation and adjust for the same set of covariates as in the conditional mean model. For comparative purposes,  we also estimate the average treatment effect parameterized by a MSMM $E[Y_{\overline a_3}]=\xi_0 +\sum_{k=0}^3 \xi_k a_k$, using IPW, iterative conditional expectation (ICE), and doubly robust estimators \citep{bang2005doubly}. The specifications of nuisance models align with those for the MSQM.

Table \ref{tab:real_data} summarizes the unadjusted, IPW, ICR, and doubly robust estimators for $\bm \theta_q$ in the MSQM at $q\in\{0.25,0.5,0.75\}$, and also includes results for $\bm\xi=[\xi_0,\xi_1,\xi_2,\xi_3]^T$ in the MSMM. For a more comprehensive view, a graphical summary of $\widehat{\bm \theta}_q$ for $q$ between 0.05 and 0.95 is provided in Web Figure S2. Without accounting for time-dependent confounding, the unadjusted approach shows that the antihypertensive treatments given at all time periods are associated with a higher SBP quantile. The remaining approaches generally suggest that the antihypertensive treatments are effective in reducing the SBP, especially at a higher quantile level. Although the overall causal conclusions from the MSMM appear similar to those from the MSQM, the MSMM fails to  capture the heterogeneity of the treatment effects over the SBP distribution. For example, the doubly robust MSQM estimator suggests that the protective effect from the third treatment $A_3$ is strengthened with a higher quantile (for example, $\widehat\theta_{3,0.25}^{\text{DR}}=-1.9$, $\widehat\theta_{3,0.5}^{\text{DR}}=-3.6$, and $\widehat\theta_{3,0.75}^{\text{DR}}=-4.5$).  This implies that patients with a higher SBP may  benefit more from the treatment, thereby avoiding harm due to excessive blood pressure. In contrast, the doubly robust MSMM estimator provides only an overall summary with $\widehat\xi_{3}^{\text{DR}}=-2.9$, suggesting that $A_3$ decreases the SBP measurements for $-2.9$ mmHg on average. 

\begin{table}[ht]
\centering
\scalebox{0.9}{
\begin{threeparttable}
\caption{Point estimation and 95\% confidence intervals of the structural parameters in analysis of the YNHHS Severe Inpatient Hypertension Dataset$^*$.\label{tab:real_data}}
\begin{tabular}{rcccc}
  \toprule
& Unadjusted & IPW & ICR/ICE$^{\dagger}$ & Doubly Robust \\ 
  \midrule
  \multicolumn{5}{l}{\textbf{A. MSQM ($q=0.25$)}} \\
  $\theta_{0.25,0}$ & 131.0 (129.4,132.6) & 134.4 (132.8,135.9) & 134.7 (133.5,136) & 134.3 (132.8,135.8) \\ 
  $\theta_{0.25,1}$ & 2.0 (0.2,3.8) & 0.9 ($-$0.9,2.7) & 0.3 ($-$1.1,1.7) & 0.7 ($-$1.1,2.4) \\ 
  $\theta_{0.25,2}$ & 0.0 ($-$1.8,1.8) & $-$2.4 ($-$4.0,$-$0.7) & $-$2.2 ($-$3.4,$-$0.9) & $-$2.4 ($-$4.0,$-$0.9) \\ 
  $\theta_{0.25,3}$ & 3.0 (1.2,4.8) & $-$1.7 ($-$3.3,$-$0.1) & $-$2.6 ($-$3.8,$-$1.4) & $-$1.9 ($-$3.4,$-$0.4) \\ 
   \midrule
  \multicolumn{5}{l}{\textbf{B. MSQM ($q=0.5$)}} \\
  $\theta_{0.5,0}$ & 150.0 (148.4,151.6) & 153.7 (152.1,155.3) & 151.6 (150.5,152.7) & 153.4 (152.0,154.9) \\ 
  $\theta_{0.5,1}$ & 1.0 $($-$0.7,2.7)$ & 0.3 ($-$1.3,1.9) & 0.0 ($-$1.3,1.2) & 0.3 ($-$1.3,1.8) \\
  $\theta_{0.5,2}$ & 0.0 $(-1.7,1.7)$ & $-$2.9 ($-$4.3,$-$1.4) & $-$1.9 ($-$3.0,$-$0.7) & $-$2.8 ($-$4.2,$-$1.4) \\
  $\theta_{0.5,3}$ & 1.0 $($-$0.7,2.7)$ & $-$3.6 ($-$5.0,$-$2.2) & $-$3.1 ($-$4.2,$-$2.1) & $-$3.6 ($-$5.0,$-$2.3) \\ 
   \midrule
  \multicolumn{5}{l}{\textbf{C. MSQM ($q=0.75$)}} \\
  $\theta_{0.75,0}$ & 167.0 (165.5,168.5) & 170.5 (168.9,172.2) & 168.5 (167.3,169.7) & 170.1 (168.6,171.7) \\
   $\theta_{0.75,1}$ & 0.0 $($-$1.7,1.7)$ & $-$0.6 ($-$2.2,1.1) & $-$0.2 ($-$1.5,1.1) & $-$0.6 ($-$2.3,1.0) \\ 
  $\theta_{0.75,2}$ & 1.0 $(-0.7,2.7)$ & $-$1.2 ($-$2.8,0.4) & $-$1.5 ($-$2.7,$-$0.3) & $-$1.0 ($-$2.6,0.5) \\  
  $\theta_{0.75,3}$ & 0.0 $(-1.7,1.7)$ & $-$4.6 ($-$6.1,$-$3.0) & $-$3.6 ($-$4.8,$-$2.5) & $-$4.5 ($-$5.9,$-$3.0) \\ 
  \midrule
  \multicolumn{5}{l}{\textbf{D. MSMM}} \\
  $\xi_{0}$ & 148.4 (147.3,149.5) & 152.1 (150.6,153.8) & 151.6 (150.6,153.0) & 151.8 (150.5,152.9) \\
   $\xi_{1}$ & 0.6 ($-0.7$,1.8) & 0.0 ($-$1.2,1.6) & 0.1 ($-1.3$,1.4) & $-$0.1 ($-$1.5,1.2) \\ 
  $\xi_{2}$ & 0.9 ($-0.4$,2.1) & $-$2.0 ($-3.3$,$-0.6$) & $-$1.8 ($-2.8$,$-0.5$) & $-$1.8 ($-$3.0,$-$0.8) \\  
  $\xi_{3}$ & 1.7 (0.4,3.0) & $-$2.9 ($-$4.2,$-$1.8) & $-$3.1 ($-4.1$,$-2.0$) & $-$2.9 ($-$4.0,$-$1.8) \\ 
   \bottomrule
\end{tabular}
\begin{tablenotes}
    \item[*] For the IPW, ICR, and doubly robust estimator of the MSQM, the 95\% confidence interval is calculated based on the derived asymptotic variance. For the IPW, ICE, and doubly robust estimator of the MSMM, the 95\% confidence interval is based on a percentile bootstrapping with 1000 replicates. The unadjusted estimators for the MSQM and MSMM are obtained by running a standard quantile regression and a linear regression with $A_1$, $A_2$, and $A_3$ as predictors, and their confidence intervals are extracted from their model-based  variance estimates.  
    \item[$\dagger$] The ICR estimator is used for the MSQM and the ICE estimator is used for the MSMM.
  \end{tablenotes} 
\end{threeparttable}} 
\end{table}

\begin{figure}
\begin{center}
\includegraphics[width=0.9\textwidth]{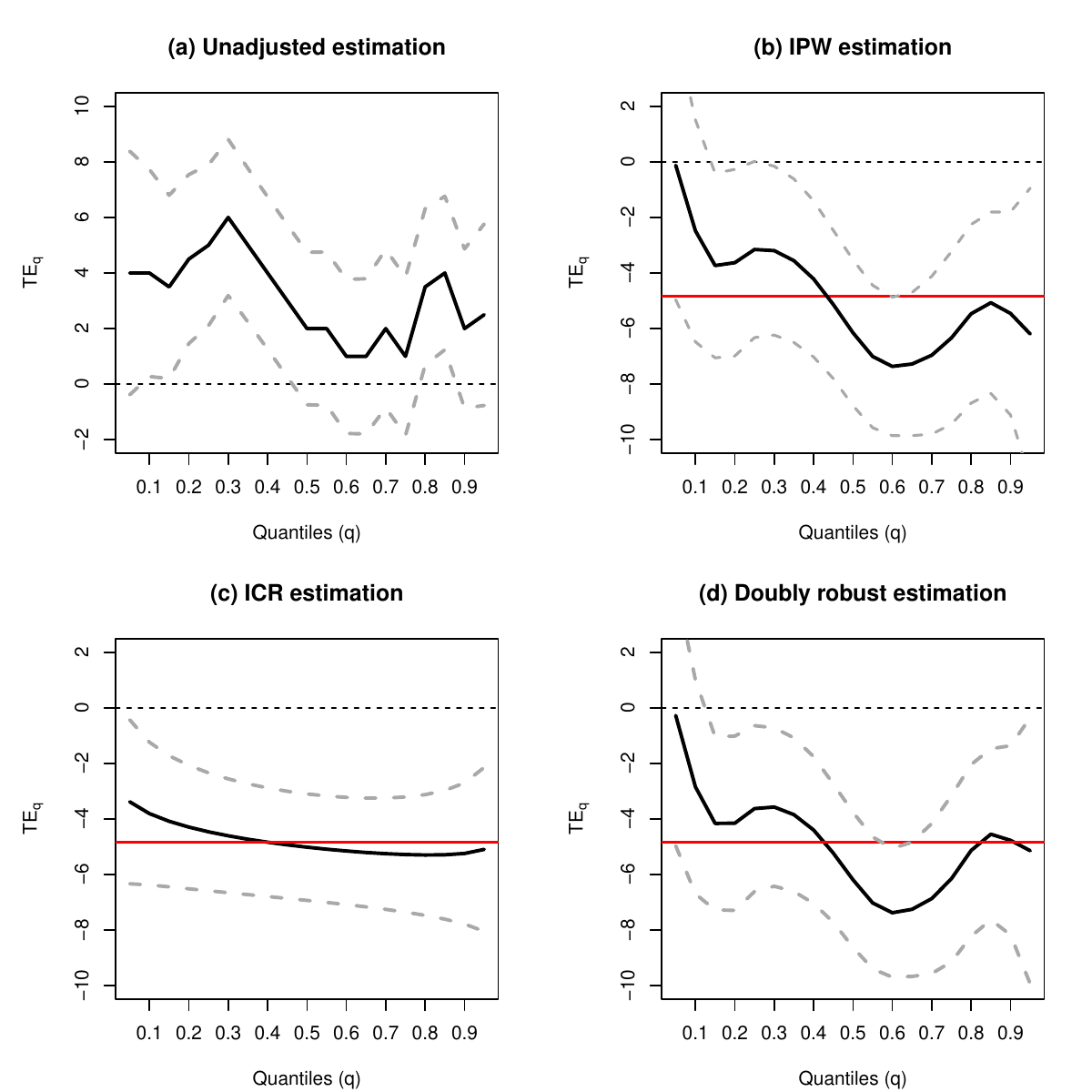}
\end{center}
\caption{ Analysis of the YNHHS Severe Inpatient Hypertension Dataset: point estimates of $\text{TE}_q$ (black line) and 95\% Wald-type confidence intervals by the derived asymptotic variance (gray dotted lines) for the total treatment effect of antihypertensives on quantiles of the patients' systolic blood pressure (SBP) at 6 hours after developing severe inpatient hypertension. The results are based on the unadjusted (panel a), IPW (panel b), ICR (panel c), and doubly robust (panel d) estimation. The red horizontal line is the doubly robust estimation of the total mean effect based on a MSMM.}
\label{fig:msqm_figure}
\end{figure}

To study the always-treat regimen, we estimate the total quantile effect by contrasting $\overline a_3=(1,1,1)$ and $\overline a_3{'}=(0,0,0)$, defined as $\text{TE}_q \equiv Q_{Y_{\overline a_3}}^{(q)} -Q_{Y_{\overline a_3{'}}}^{(q)}=\sum_{k=1}^3\theta_{q,k}$. For comparison, the total mean effect ($\text{TE}_{\text{mean}}$) is also obtained from the doubly robust MSMM estimator. Figure \ref{fig:msqm_figure}  reports $\widehat{\text{TE}}_q$ and $\widehat{\text{TE}}_{\text{mean}}$ for $q$ ranging from 0.05 to 0.95. Except for the unadjusted approach, all IPW, ICR, and doubly robust estimators suggest a negative total quantile effect. The point estimates of IPW and doubly robust estimators are more closely aligned, but the doubly robust estimator provides slightly narrower confidence intervals. As $q$ increases, both the IPW and doubly robust estimators indicate that $\text{TE}_q$ first strengthens up to $q=0.6$ and then gradually declines. For example, $\text{TE}_{q}$ obtained from the doubly robust MSQM increases from $-$0.3 (95\% CI: [$-$4.9,4.4]) at $q=0.05$ to $-$7.4 (95\% CI: [$-$9.7,$-$5.1]) at $q=0.6$, and then slightly reduces to $-$5.2 (95\% CI: [$-$9.9,$-$0.3]) at $q=0.95$. The ICR estimator exhibits a smoother and flatter pattern, suggesting that $\text{TE}_{q}$ first increases from $-$3.4 (95\% CI: [$-$6.3,$-$0.4]) at $q=0.05$ to $-$5.3 (95\% CI: [$-$7.5,$-$3.1]) at $q=0.8$, and then slightly decreases to $-$5.1 (95\% CI: [$-$8.0,$-$2.1]) at $q=0.95$. Although the patterns of the total effect estimates are similar across methods, the smoothness and flatness under ICR in this data example are mainly driven by its full distributional assumptions. 

\subsection{Sensitivity analysis to unmeasured confounding} 

We further evaluate the sensitivity of $\widehat{\text{TE}}_{0.5}$ obtained by the doubly robust approach to departure from sequential ignorability. We use the working confounding function \eqref{sf}, and 
consider the following functional forms of the \textit{peak} component:
$$
r_k(\overline a_3, \overline{\bm l}_k;\bm\gamma_{k1}) = \gamma_{k1}(2 a_k-1), \quad \text{for } k=1,2,3, 
$$
The sensitivity parameter $\gamma_{k1}$ was selected between  $[-0.5,0.5]$ such that total variation distance between the two CDFs in \eqref{sf_true} does not exceed 0.5. When $\gamma_{k1}>0$, we have $c_k(y,\underline{a}_{k+1},1,\overline{a}_{k-1}, \overline{\bm l}_k;\bm\gamma_{k1})>0$ but $c_k(y,\underline{a}_{k+1},0,\overline{a}_{k-1},  \overline{\bm l}_k;\bm\gamma_{k1})<0$, suggesting that, conditional on $\overline{\bm L}_k=\overline{\bm l}_k$ and $\overline A_{k-1}=\overline a_{k-1}$, the  treated individuals in the $k$-{th} period tends to have a lower potential SBP value (under all possible $\overline{a}_3\in \overline{\mathbb A}_3$) than the untreated individuals; in other words, healthier patients were more likely to be treated in the $k$-{th} period. Otherwise, $\gamma_{k1}<0$ suggests that unhealthier patients were more likely to be treated in the $k$-{th} period. To measure the area under the working confounding function curve, we set the \textit{area} component, $b_k(\overline a_K, \overline{\bm l}_k;\bm\gamma_{k2})$, to a constant value $\gamma_{k2}\in [2,6]$ such that the $1-$Wasserstein distance between the two CDFs used in the confounding function does not exceed 6. If we assume that there is only a mean shift between the two CDFs in \eqref{sf_true}, $\gamma_{k2}$ also approximately equals to the absolute mean difference in the potential SBP values between the treated and untreated individuals in the $k$-{th} period, conditional on $\overline{\bm L}_k=\overline{\bm l}_k$ and $\overline A_{k-1}=\overline a_{k-1}$. By Proposition 2 in the Web Appendix G, the \textit{location} component, $m_k(\overline a_3,\overline{\bm l}_k;\bm\gamma_{k3})$, is closely connected to the conditional mean of potential SBP given $\overline{\bm L}_k=\overline{\bm l}_k$ and $\overline A_{k}=\overline a_{k}$, i.e., $\bm\delta_k^T \bm g_k(\overline a_3,\overline{\bm l}_k)$. We exploit this connection and set $m_k(\overline a_K,\overline{\bm l}_k;\bm\gamma_{k3}) = \gamma_{k3}\Big(\widehat{\bm\delta}_k^T \bm g_k(\overline a_K,\overline{\bm l}_k)\Big)$ such that the difference between the two CDFs used in the confounding function \eqref{sf_true} is maximized at $\gamma_{k3}$ times the estimated mean of the potential SBP conditional on $\overline{\bm L}_k=\overline{\bm l}_k$ and $\overline A_{k}=\overline a_{k}$. For this specification,  $\widehat{\bm\delta}_k$ is obtained from the ICR approach and $\gamma_{k3}$ is varied within $[0.9,1.1]$.

\begin{figure}
\begin{center}
\includegraphics[width=1\textwidth]{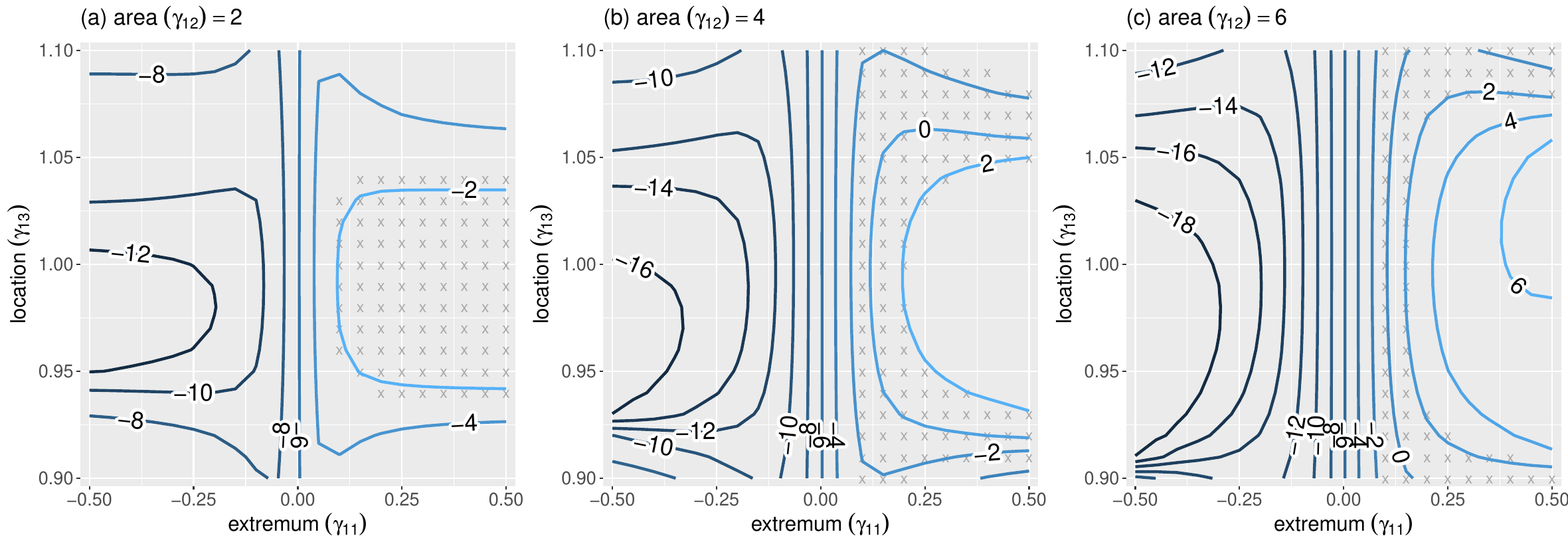}
\end{center}
\caption{Sensitivity analysis for $\text{TE}_{0.5}$ to unmeasured baseline confounding, contours of the bias-corrected doubly robust estimates for fixed values of $(\gamma_{11},\gamma_{13})$, when $\gamma_{12}=2$ (Panel a), 4 (Panel b), 6 (Panel c). Here, the region with `$\times$' indicates $(\gamma_{11},\gamma_{13})$ such that the corresponding 95\% interval estimate covers 0.}
\label{fig:exp2_sa1}
\end{figure}

To disentangle effects of unmeasured confounding at different periods, we report $\widehat{\text{TE}}_{0.5}$ to unmeasured confounding at $k=1,2,3$, separately; i.e., when we study unmeasured confounding at period $k$,  confounding functions at other periods are fixed to $0$. The effect estimate $\widehat{\text{TE}}_{0.5}$ under sequential ignorability is $-6.2$ ($95\%$ CI: $[-8.6,-3.8]$). Figure \ref{fig:exp2_sa1} presents the sensitivity results to unmeasured baseline confounding. We observed that 
a larger positive value of $\gamma_{11}>0$ moves $\widehat{\text{TE}}_{0.5}$ towards null, while a larger negative value of $\gamma_{11}<0$ corresponds to a stronger negative $\widehat{\text{TE}}_{0.5}$. Additionally, $\widehat{\text{TE}}_{0.5}$ is more sensitive to larger values of $\gamma_{12}$, as evidenced by the denser contours from panel (a) to panel (c). This is somewhat expected since larger values of $\gamma_{11}$ and $\gamma_{12}$ indicate a greater magnitude of unmeasured baseline confounding. Finally, $\widehat{\text{TE}}_{0.5}$ presents non-monotonic behavior with respect to $\gamma_{13}$. Specifically, $\widehat{\text{TE}}_{0.5}$ is more sensitive when $\gamma_{13}$ is closer to 1 but relatively more robust when $\gamma_{13}$ moves away from 1. For example, when fixing $\gamma_{11}=0.25$ and $\gamma_{12}=4$, $\widehat{\text{TE}}_{0.5}$ first increases from $-2.8$ (95\% CI: $[-5.6,-0.1]$) to 2.7 (95\% CI: $[0.7,4.6]$) as $\gamma_{13}$ moves from 0.9 to 1 and then drops to $-2.5$ (95\% CI: $[-5.0,0.1]$) as $\gamma_{13}$ moves toward 1.1. Of note, $\widehat{\text{TE}}_{0.5}$ remains negative (indicating a positive effect of antihypertensive treatment) when either $\gamma_{11}\leq 0.1$ or $\gamma_{13} \leq 2$ regardless of the value of $\gamma_{12}$. Thus, our conclusion under sequential ignorability is not overturned if the total variation or the 1-Wasserstein distance between the two CDFs in the confounding function is less than 0.1 or 2, respectively. Sensitivity results to unmeasured confounding at periods 2 and 3 are given in Web Figures S3--S4. The influence of unmeasured confounding at period 2 and 3 to $\widehat{\text{TE}}_{0.5}$ is smaller as compared to the unmeasured baseline confounding, assuming the same value for $\bm\gamma_k$. 

\section{Discussion}
\label{sec:disc}
{\color{black}Our proposed doubly robust approach allows for misspecifications of the propensity score or conditional outcome regression models, but it still relies on a parametric MSQM that correctly represents the relationship between treatment regimens and potential outcome quantiles. If the MSQM is misspecified, there is a risk of obtaining treatment effect estimates that are much less interpretable. 
When multiple candidate specifications for the MSQM are viable, it is of interest to establish appropriate  information criteria for model selection. Similar to \cite{platt2013information}, a quasi-information criterion for the IPW estimation of MSQM can be constructed as $\text{QIC} = 2p+2n Q_{loss}(\widehat{\bm\alpha},\widehat{\bm \theta}_q^{\text{IPW}})$, where $2p$ is a penalty term for the number of parameters used in the MSQM and $Q_{loss}(\widehat{\bm\alpha},\widehat{\bm \theta}_q^{\text{IPW}})= \mathbb{P}_n\left[\frac{\rho(\overline A_K,\bm Z)}{ \overline{\pi}_K(\overline A_K, \overline{\bm L}_K;\widehat{\bm\alpha})} \delta_q\left(Y - h(\overline A_K, \bm Z;\widehat{\bm \theta}_q^{\text{IPW}})\right)\right]$ is an inverse probability weighted empirical quantile loss function which converges to the true quantile loss function defined in Web Appendix A. Because a smaller value of the quantile loss function suggests a better approximation of MSQM to the true quantile causal curve, a small QIC reflects a better MSQM fit to the underlying data. Although defining the quasi-information criterion for the IPW estimator is relatively straightforward, similar criteria for the ICR and doubly robust estimators are not yet available, signifying future developments.}

\section*{Acknowledgement}
Research in this article was supported by the Patient-Centered Outcomes Research Institute\textsuperscript{\textregistered} (PCORI\textsuperscript{\textregistered} Award ME-2021C2-23685). The statements presented in this article are solely the responsibility of the authors and do not necessarily represent the views of PCORI\textsuperscript{\textregistered}, its Board of Governors or Methodology Committee.  
Data used in Section \ref{sec:examples} was obtained with the assistance and supervision of the Clinical and Translational Accelerator (CTRA), Department of Medicine, Yale School of Medicine and the Joint Data Analytics Team.

\section*{Supplementary material}
Web Appendices, Tables, Figures, and R code for simulations can be found in the GitHub repository: \url{https://github.com/chaochengstat/MSQM}.




\bibliographystyle{biom}
\bibliography{paper-ref}






\label{lastpage}

\end{document}